\documentclass{WileyMSP-template}
\usepackage{mathrsfs}
\usepackage[dvipsnames]{xcolor}

\usepackage{amsmath}
\usepackage{bbm}

\usepackage[colorlinks=true,
            linkcolor=blue,    
            citecolor=blue,      
            urlcolor=blue         
           ]{hyperref}

\begin{document}

\pagestyle{fancy}
\rhead{\includegraphics[width=2.5cm]{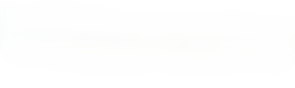}}

\title{Exceptional Antimodes in Multi-Drive Cavity Magnonics}

\maketitle


\author{Mawgan A. Smith$~^{1,*}$}
\author{Ryan D. McKenzie$~^{2,*}$}
\author{Alban Joseph$~^1$}
\author{Robert L. Stamps$~^2$}
\author{Rair Mac\^edo$~^1$}


\begin{affiliations}

$^1$James Watt School of Engineering, Electronics \& Nanoscale Engineering Division, University of Glasgow, Glasgow G12 8QQ, United Kingdom\\

$^2$Department of Physics and Astronomy, University of Manitoba, 30A Sifton Road, Winnipeg, R3T 2N2, Manitoba, Canada

$^*$\emph{These authors contributed equally to this work.}

\end{affiliations}




\begin{abstract}
Driven-dissipative systems provide a natural setting for the emergence of exceptional points---i.e. non-Hermitian degeneracies where eigenmodes coalesce.
These points are important for applications such as sensing, where enhanced sensitivity is required, and exhibit interesting and useful phenomena that can be controlled with experimentally accessible parameters.
In this regard a four-port, three-mode, cavity-magnonics platform is demonstrated in which two microwave excitations can be precisely phase shifted and/or attenuated relative to one another. 
Destructive interference between the hybridised cavity-magnon modes is shown to give rise to antimodes (antiresonances) in the transmission spectrum, enabling coherent perfect extinction of the outgoing signals at selected ports. 
This interference can be used to actively tune the position and properties of exceptional points, without the fine tuning conventionally required to obtain exceptional points.
Such controllable, interference-based engineering of exceptional points provides a practical and flexible pathway toward next-generation, high-sensitivity sensing devices operating at microwave frequencies.

\end{abstract}


\section{Introduction}

Cavity magnonics is based on coherent coupling between magnons and photons in resonator systems and has generated a wealth of interesting phenomena and potential applications \cite{FlebusReview}. 
Recent examples include precision magnetometry for quantum-enhanced detection of single magnons in hybrid quantum systems \cite{LQ, LQReview, LQScience, Crescini}, and measurement of a soft mode associated with the ferromagnetic to paramagnetic quantum phase transition in a quantum-Ising magnet \cite{LiberskySM, StampGSM}. 
The future for cavity magnonics is wide ranging and exciting with possible applications ranging from emerging technologies such as microwave-to-optical transduction \cite{Hisatomi, Engelhardt, Puel} to dark matter detection \cite{Barbieri, CresciniDM, FlowerDM}. 

Cavity magnon-polaritons are hybridized eigenstates of a system in which there is coherent energy exchange between magnons and photons which is a hallmark of Hermitian systems \cite{Imamoglu, Soykal, Huebl, Zhang}. 
An experimental signature of these quasiparticles is level repulsion which is a splitting of otherwise degenerate normal modes due to the coupling, common to a wide range of coupled systems.  
On the other hand, a different phenomenon termed 'level attraction' has also been demonstrated in cavity magnonics.
Level attraction was originally attributed to the coupling between magnons and photons acquiring a dissipative character, either through a common resevoir or auxillary mode \cite{Kubala, Heiss2004, Eleuch, Dhara2017, HarderPRL, Harder2021, Bernier2014, Bernier, Yu, Proskurin2018, Proskurin2019, Proskurin2021,  Bhoi, Yang2019, Yang2020, Xu, WangNonrecip, Yao, YuDissipativeCoupling, RaoTravelingPhotons, WangHu, Xiao, Wang, Lu, Hong, RaoAntires, Hao}. 
This phenomenon is a signature of non-Hermitian systems \cite{Ashida}.
Adjacent work on controlling cavity-magnon systems using a two-tone excitation also observed features of level attraction in the spectral response \cite{Boventer1,Boventer2}.
Although both two-tone excitation  \cite{grigoryan, Boventer1, Boventer2} and anti-resonance \cite{RaoAntires} experiments modelled level attraction using dissipative coupling, it was recently shown that such spectral response is observable with only coherent coupling in the system; demonstrating this to be an interference-based phenomena \cite{rao2021interferometric, Bourcin, gardin2025level}.

A transition from level repulsion to level attraction spectra can be affected by tuning a system parameter such as the coupling strength through a degeneracy point of the uncoupled modes.
For dissipative coupling mechanisms, it is now well understood that the point at which the real part of the two modes coalesce is known as an exceptional point. 
At this point the modes are degenerate and the eigenstates merge \cite{Berry, Heiss2004, Heiss, zhang2017observation}. These are spectral singularities topological structure and exhibit unique properties, including topological mode switching \cite{Heiss1999, Dembowski2001, Dembowski2004, Hernandez, Lefebvre, Uzdin, Milburn, xu2016topological, Ghosh, Doppler, Schumer}, and braiding of the complex eigenvalues \cite{Wojcik, Patil, Chavva}. These phenomena have recently been measured and studied by several different groups \cite{rao2024braiding, LambertModeSwitch, ZhangModeSwitch}. Additionally, exceptional points with a nontrivial spectral response that amplifies small perturbations, leading to sensitivity that scales nonlinearly with perturbation strength are  promising as high-sensitivity detectors \cite{Wiersig2014, Wiersig2016, chen2017exceptional, hodaei2017enhanced, CaoSensing, Zhong, chen2019, ZhangSensing, WiersigReview, LiStochasticEP}.

In this work, we present a two-dimensional cavity magnonics device engineered to support interference-based exceptional points using a multi-drive architecture.
In contrast to more conventional resonant exceptional points that arise from poles of the scattering ($S$-) matrix, interference-based exceptional points are associated with \emph{zeros} of the S-matrix \cite{WangCPAEP}. This leads to an interesting distinction that we explore in this work, namely the difference between coherent perfect absorption (CPA) and coherent perfect extinction (CPE).

In general, the zeros of the S-matrix are complex--if a zero of the S-matrix occurs at a real frequency there is CPA where for a specific input configuration in a multiport system, the outgoing signal at all ports vanishes through destructive interference\cite{chongCPA, Baranov}. The less restrictive general condition is CPE and can be achieved by considering a submatrix $\widetilde{S}$ of the S-matrix \cite{guo2023singular}. Extinction occurs at one or more selected ports while transmission through the remaining channels persist. 

The CPE effect encompasses many observed scattering phenomena. As an example, reflectionless scattering modes indicate frequencies at which reflection is fully suppressed at complex zeros of det$[\widetilde{S}]$ \cite{SweeneyRSM}. 
This is a clear and robust experimental signature in transmission and reflection spectra and are topologically nontrivial \cite{guo2023singular}. 

In the multiport system discussed here, interference control is enabled by multiple excitation fields that can be adjusted to create CPE and antiresonant exceptional points without the need for fine tuning of intrinsic system parameters. This makes interference-based exceptional points experimentally accessible and robust, offering new opportunities for sensitive detection, signal routing, and readily controllable non-Hermitian physics in cavity magnonics.

In Sec. \ref{sec:system} of this paper, we describe our four port, three mode, cavity-magnonics system and demonstrate novel scattering phenomena including CPA and CPE. Transmission through the microwave network is determined by the $S$-matrix, which we consider in Sec. \ref{sec:Smatrix}.
Furthermore, we use an input-output formalism adopted from quantum optics to determine the $S$-matrix with details of the calculation provided in the Supplemental Information. 
In Sec. \ref{sec:Eres} we characterise the resonators in the absence of  magnetic material. In Sec. \ref{sec:TTD}, we detail the advantages of multiple excitation fields, and in Sec. \ref{sec:EAM} we show how full phase and amplitude control over the excitations can be introduced to obtain novel, interference-based, exceptional points for any given system parameters.

\section{The System}
\label{sec:system}

\begin{figure}[bt]
\centering
\includegraphics[width=0.4\linewidth]{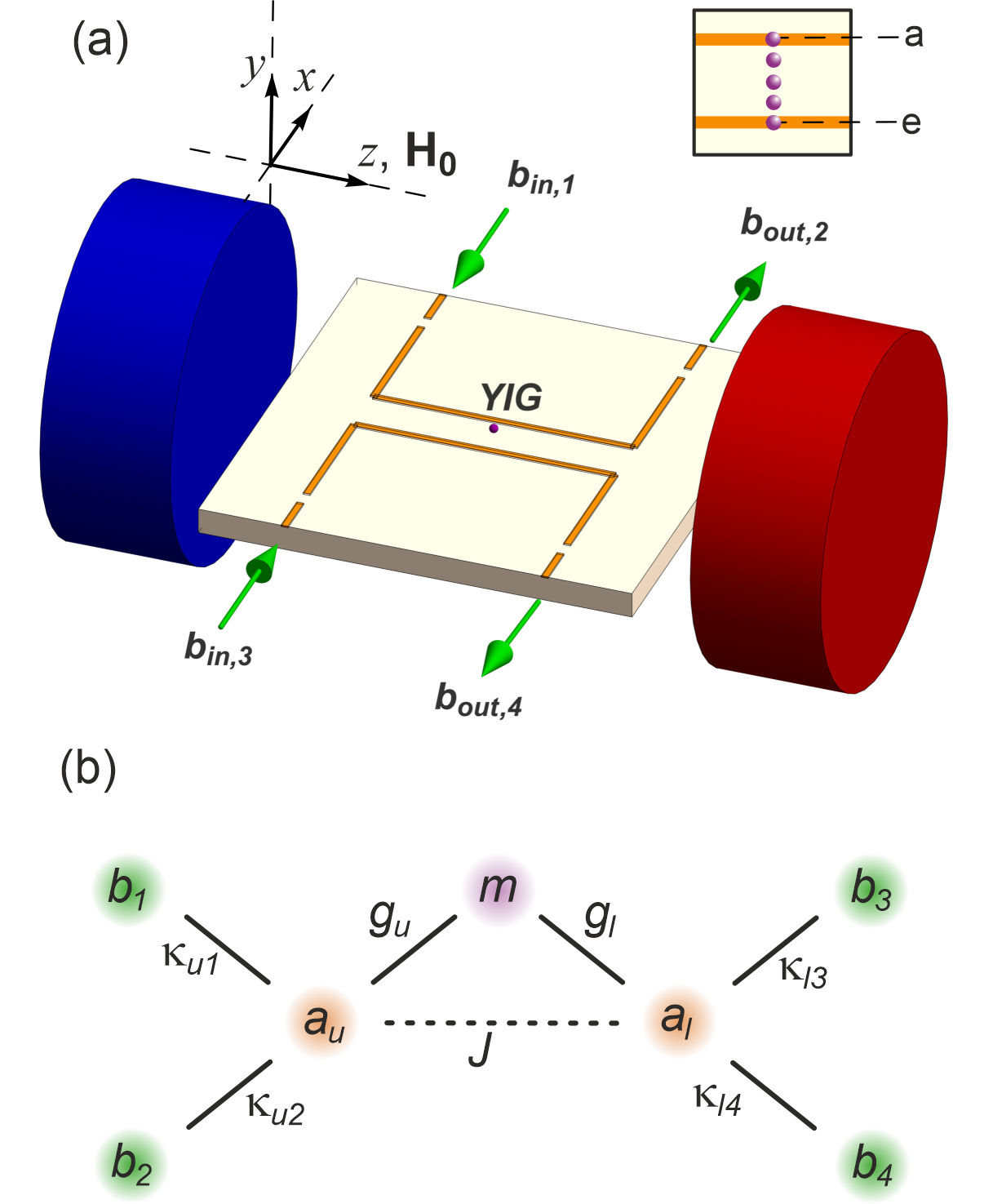}
\caption{(a) Schematic of the experimental platform in the input–output configuration used for the multi-excitation measurements.  
Two open transmission line resonators are placed in close proximity, allowing inductive coupling between their modes with strength $J$.  
A YIG sphere is positioned such that it can couple to both resonators. 
Independent microwave drives can be applied at the four ports $b_{\mathrm{in},1\text{--}4}$, with corresponding output fields $b_{\mathrm{out},1\text{--}4}$. 
(b) Interaction diagram of the system Hamiltonian. 
The upper ($a_u$) and lower ($a_l$) resonator modes are coherently coupled with strength $J$, and each couples to the common magnon mode ($m$) with coupling strengths $g_u$ and $g_l$, respectively. 
Each resonator is also coupled to two independent input–output channels $b_i$.}
  \label{fig:combined1}
\end{figure}

Schematics of our four-port, three-mode, cavity-magnonics system is shown in Fig.~\ref{fig:combined1}. 
The experimental configuration is shown in (a) and consists of two transmission line resonators wherein independent microwave excitations can be applied at the four ports $b_{\mathrm{in},1\text{--}4}$, with corresponding output fields $b_{\mathrm{out},1\text{--}4}$.
Here. we will primarily focus on the case when input excitations are applied to ports $b_{\mathrm{in},1}$ and $b_{\mathrm{in},3}$ and measured at port $b_{\mathrm{out},2}$ (or $b_{\mathrm{out},4}$). 
At the centre, the two transmission lines are positioned close enough to enable inductive coupling with strength $J$, allowing controllable interference between the two drive pathways by controlling the amplitude and phase of both $b_{\mathrm{in},1}$ and $b_{\mathrm{in},3}$.
An yttrium–iron–garnet (YIG) sphere is positioned close to the transmission lines, and its location can be controlled from directly between them to directly above either line, as shown in the inset of Fig.~\ref{fig:combined1}(a), allowing control over the relative coupling strengths between each resonator and magnon modes in the YIG sphere.
A static magnetic field is applied externally along $z$, that is, parallel to the transmission lines (at the sample position).  

This system is modelled as a pair of degenerate modes, that we call the upper (lower) bosonic annihilation operator $a_u$ ($a_l$) \cite{hsiehOpenrings}. 
We can then associate $a_u$ to the transmission line resonator excited by $b_{\mathrm{in},1}$ and $a_l$ to the resonator excited by $b_{\mathrm{in},3}$.
Each resonator mode is coupled to a common magnon mode, the Kittel mode in YIG, with annihilation operator $m$, and coupling strengths $g_u$ and $g_l$. 
Treating all interactions in the rotating-wave approximation, the system Hamiltonian can be written as
\begin{align}
\frac{\mathcal{H}_{\text{sys}}}{\hbar} = \omega_a &a_u^{\dag} a_u + \omega_a a_l^{\dag} a_l
+ \omega_m m^{\dag} m
+[J a_u^{\dag} a_l + g_u a_u^{\dag} m + g_l a_l^{\dag} m + H.c.]
\end{align}
The frequency of the magnon mode can be tuned with the applied static magnetic field $\mathbf{H_0}$ through $\omega_m = \gamma \mu_0 H_0$, where $\gamma/2\pi$ = 28 GHz/T is the gyromagnetic ratio for YIG and $\mu_{0}$ is the permeability of free space.

The upper and lower modes are coupled to input-output ports on their two sides, modelled by four independent baths.
The total Hamiltonian is then $\mathcal{H}=\mathcal{H}_{\text{sys}}+\mathcal{H}_{\text{bath}}+\mathcal{H}_{\text{int}}$, where, treating the bath modes in the continuum limit, one has  
\begin{align}
\label{eq:Hbath}
\mathcal{H}_{\text{bath}} = \hbar \int d\omega \sum_{n=1}^4 \omega b_n^{\dag}(\omega) b_n(\omega).
\end{align}
The system-bath interaction can be written as
\begin{align}
\mathcal{H}_{\text{int}} = i \hbar \int d\omega \ \boldsymbol{b}^{\dag} K_0^{\dag} \boldsymbol{a}_{0} + H.c.,
\end{align}
where the bath and cavity vectors are $\boldsymbol{b}^T = [b_1,b_2,b_3,b_4]$ and $\boldsymbol{a}_0^T = [a_u,a_l]$, and the coupling matrix is
\begin{align}
\label{eq:K0}
K_0 = \left[\begin{array}{cccc}
\kappa_{u1} & \kappa_{u2} & 0 & 0 \\
0 & 0 & \kappa_{l3} & \kappa_{l4} \end{array}\right].
\end{align}
The couplings are assumed to be frequency independent (Markov approximation \cite{GardinerZollerBook, WallsMilburn}) and real. The reality of the coupling constants neglects any phase shift of the input/output that may occur at the resonator ports. A schematic of the coupled mode interactions is given in Fig. \ref{fig:combined1}(b).
For further details of the experimental setup, and discussion of the phase shifts neglected in the coupled mode theory, see Supporting Information~\ref{ap:methods}.

The independent resonators are degenerate; the inductive coupling between the resonators splits the modes into a low-frequency and high-frequency mode. It is convenient to work in a basis, referred to here as the cavity basis, in which the resonator component of the Hamiltonian is diagonal. 
In the cavity basis, the system Hamiltonian can be written as
\begin{align}
\label{eq:Hsys}
\frac{\mathcal{H}_{\text{sys}}}{\hbar} = \omega_1 &a_1^{\dag} a_1 + \omega_2 a_2^{\dag} a_2
+ \omega_m m^{\dag} m
 +[g_1 a_1^{\dag} m + g_2 a_2^{\dag} m + H.c.]
\end{align}
where $\omega_1=\omega_a-J$ and $\omega_2=\omega_a+J$. 

The cavity operators 
\begin{align}
a_1=\frac{1}{\sqrt{2}} (a_u-a_l) \  \ \text{and}\ \  a_2=\frac{1}{\sqrt{2}} (a_u+a_l),
\end{align}
are antisymmetric and symmetric combinations of the system operators and the couplings are $g_1=(g_u-g_l)/\sqrt{2}$ and $g_2=(g_u+g_l)/\sqrt{2}$. 
The coherent couplings between the cavity photons and magnons lead to the formation of magnon-polaritons. 
Note that when the YIG sphere is centred between the resonators ($g_u=g_l$), the low-frequency cavity mode is decoupled from the magnons. 
With the YIG sphere in an asymmetric position, one has indirect coupling between the cavity photons via the magnons \cite{hyde2016indirect}.

The system-bath interaction must also be rotated to the cavity basis. One finds
\begin{align}
\label{eq:Hint}
\mathcal{H}_{\text{int}} = i \hbar \int d\omega \ 
\boldsymbol{b}^{\dag} K^{\dag} \boldsymbol{a} + H.c.
\end{align}
where $\textbf{a}^T=[a_1,a_2]$ and the coupling matrix is
\begin{align}
\label{eq:couplings}
K = \frac{1}{\sqrt{2}} \left[\begin{array}{cccc}
\kappa_{u1} & \kappa_{u2} & -\kappa_{l3} & -\kappa_{l4} \\
\kappa_{u1} & \kappa_{u2} & \kappa_{l3} & \kappa_{l4} \end{array}\right].
\end{align}
The cavity photons share a common coupling to each of the baths. 
This can lead to a purely dissipative coupling between the cavity photons \cite{MetelmannClerk, WangHu}, provided there is asymmetry in the system-bath couplings. 
In this paper, for the most part, we will assume that the system bath couplings are equal $\kappa_{un}=\kappa_{ln}=\kappa$, in which case this dissipative coupling mechanism is absent. Further details are provided in Supporting Information~\ref{ap:IO}.

Transmission and reflection are determined by the S-matrix, which we now consider.

\subsection{The S-matrix}
\label{sec:Smatrix}

For a given input ${x}^{OUT}$, the $S$-matrix determines the output, ${x}^{IN}$, from a microwave network by
\begin{align}
\boldsymbol{x}^{OUT} = S \boldsymbol{x}^{IN}.   
\end{align}
In the four-port system introduced earlier, the $S$-matrix is a $4 \times 4$ matrix whose elements determine the output at the $m^{th}$ port for a given input at the $n^{th}$ port, $x_m^{OUT} = S_{mn} x_n^{IN}$. 
The $S$-matrix can be determined by using the input-output formalism \cite{GardinerCollett, GardinerZollerBook, WallsMilburn} and since the method is now well-established, we relegate it to Supporting Information~\ref{ap:IO} and here, we focus on the result.

With equal system-bath couplings, there are several distinct transmission pathways. 
These are direct transmission through the upper and lower resonators, $t_{du}$ and $t_{dl}$, and indirect transmission $t_i$, in which the input from one resonator is transmitted out of the second. 
Reflection is related to direct transmission by $r_{u,l}=-1-t_{du,l}$ and symmetry requires reciprocal processes to be equivalent. 
Therefore, the full $S$-matrix is
\begin{align}
\label{eq:Smatrix}
S = \left[ \begin{array}{cccc}
r_u & t_{du} & t_i & -t_i  \\
t_{du} & r_u & -t_i & t_i  \\
t_i & -t_i & r_l & t_{dl}  \\
-t_i & t_i & t_{dl} & r_l  \\
\end{array} \right],
\end{align}
where the minus sign on certain indirect transmission matrix elements results from the phase convention chosen for the inputs and outputs \cite{WallsMilburn} (see Eq. \ref{eq:x} of Supporting Information~\ref{ap:IO}).

With all couplings set to $\kappa$, direct transmission is given by 
\begin{align}
\label{eq:Td}
t_{du;dl} = -i \pi \kappa^2
\biggr[\frac{S_{1m}+S_{2m} \pm 2g_{12}S_{1m}S_{2m}}
{1-g_{12}^2S_{1m}S_{2m}}\biggr],
\end{align}
where $t_{du}$ ($t_{dl}$) has the plus (minus) sign, and indirect transmission is given by
\begin{align}
\label{eq:Ti}
t_i = i \pi \kappa^2
\biggr[\frac{S_{1m}-S_{2m}}{1-g_{12}^2S_{1m}S_{2m}}\biggr],
\end{align}
where
\begin{align}
S_{jm}^{-1} = \omega-\omega_j
+i\frac{\gamma_j}{2}-\frac{g_j^2}{\omega-\omega_m+i\gamma_m/2},
\end{align}
and
\begin{align}
g_{12}=\frac{g_1g_2}{\omega-\omega_m+i\gamma_m/2}.
\end{align}
The poles of $S_{jm}$ are the low ($j=1$) and high ($j=2$) frequency magnon-polariton modes, and $g_{12}$ is an indirect coupling mediated by the magnons. 

To better understand the transmission matrix elements, we consider the system when the YIG sphere is centred between the resonators [as per experimental setup depicted in Fig.~\ref{fig:combined1}(a)]. 
In this case, the couplings to the upper and lower resonators are equal, $g_u=g_l$, and 
\begin{align}
g_{12} = \frac{g_u^2-g_l^2}{2(\omega-\omega_m+i\gamma_m/2)}  = 0.
\end{align}
The direct transmission matrix elements are then equal, and Eqs. (\ref{eq:Td}) and (\ref{eq:Ti}) indicate that there is direct and indirect transmission at the magnon-polariton mode frequencies. Corrections to this simple result arise due to asymmetry in the placement of the YIG sphere. 

The form of the transmission matrix elements given in Eqs. (\ref{eq:Td}) and (\ref{eq:Ti}) help clarify the physics of transmission through the resonator network. From a more practical standpoint, it is better to write the transmission matrix elements as a product of their poles and zeros, or a sum over their poles with complex spectral weights. Scattering phenomena such as coherent perfect absorption and extinction are closely tied to properties of the S-matrix, which we consider next.

To better understand coherent perfect absorption (see Refs. \cite{chongCPA, Baranov}) we define the following inputs and outputs: $\boldsymbol{y}^{IN/OUT} = K\boldsymbol{b}^{IN/OUT}$ (the relation between $\boldsymbol{b}^{IN/OUT}$ and $\boldsymbol{x}^{IN/OUT}$ is given in Eq. (\ref{eq:x}) of Supporting Information~\ref{ap:IO}). 
These inputs and outputs are related by a $2 \times 2$ $S$-matrix, $\boldsymbol{y}^{OUT} = S_{\Omega}\boldsymbol{y}^{IN}$, where (see Eq. (\ref{eq:Somega}) of Supporting Information~\ref{ap:IO}) 
\begin{align}
S_{\Omega}=(\Omega-i2\pi K K^{\dag}) \Omega^{-1}.
\end{align}
The condition for coherent perfect absorption is $\text{det}[S_{\Omega}]=0$. The zeros of the $S$-matrix are determined by $\text{det}[\Omega-i2\pi K K^{\dag}]=0$. Generally, these zeros are complex. Coherent perfect absorption occurs when a zero occurs at a real frequency. The modes of the damped system follow from the zeros of $\text{det}[\Omega]$. If two eigenvectors of $\Omega$ coalesce, one has a resonant exceptional point. If eigenvectors of $\Omega-i2\pi K K^{\dag}$ coalesce, one has an absorbing exceptional point, or a coherent perfect absorbing exceptional point if the eigenvalues are real. 

Of primary interest here are multi-drive systems in which there are two inputs, $x_m^{IN} = \delta e^{i\Phi} x_n^{IN}$. If one considers output at port $p$, one has a $1 \times 2$ submatrix of the full S-matrix. In general, if there are fewer output ports than input ports, coherent perfect extinction is possible \cite{guo2023singular}. In this particular example, there is coherent perfect extinction when $x_p^{OUT} = 0 = (S_{pn} + \delta e^{i\Phi} S_{pm}) x_n^{IN}$. With two drive fields, one may achieve this by setting $\delta e^{i\Phi} = -S_{pn}/S_{pm}$.

Prior to our analysis of the full cavity-magnonics system, we characterise the empty resonators in Sec. \ref{sec:Eres}. 

\subsection{Resonator Characterisation}
\label{sec:Eres}
The two coupled resonators of the four port system is a two-mode system which, when loaded with YIG becomes a three-mode system which can be tuned by an external magnetic field through the magnon mode. Before discussing the YIG loaded system, we discuss features of the empty two-mode system such that it is not affected by a magnetic field.

\begin{figure}[!h]
	\centering
	\mbox{\includegraphics[width=8.2cm]{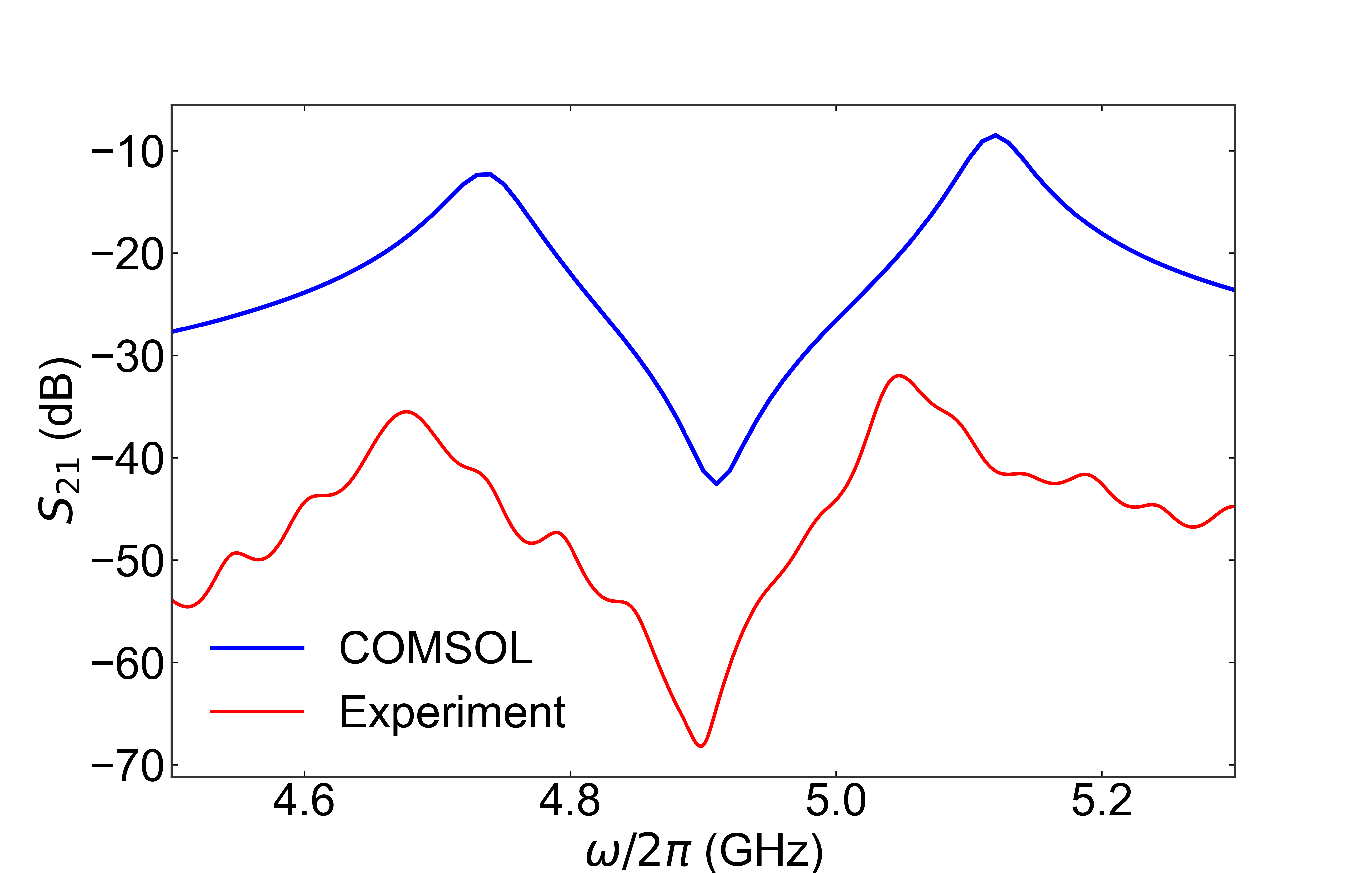}}        
    \caption{$S_{21}$, corresponding to the direct transmission in Eq. (\ref{eq:tdti}), of the empty resonator. The transmission in the experiment is significantly less than the simulation primarily because of losses due to the external circuitry connected to the VNA. There is also a static oscillating background due to interference in the cables that is not present in the simulation. In this case it does not greatly obscure fitting to the input-output model.}
    \label{fig:Eres}
\end{figure}

In Fig.~\ref{fig:Eres}, we present the experimentally measured transmission spectra (in decibels, $10 \log_{10}|S_{n1}|^2$) when Port 1 ($b_{\text{in},1}$) is driven in the special case of no YIG (empty resonators). 
While the analytical framework discussed earlier will be used to describe the coupled system, numerical modelling offers a complementary and more comprehensive approach. Specifically, we employ COMSOL Multiphysics\textregistered~\cite{comsol} to perform full-wave electromagnetic simulations of the complete resonator geometry shown in Fig.~\ref{fig:combined1}(a). 
Unlike the analytical model, which necessarily relies on simplifications such as ideal coupling, the COMSOL simulation captures the full spatial and material complexity of the structure, including boundary conditions and field distributions. 
The resulting simulated spectrum, also shown in Fig.~\ref{fig:Eres}, demonstrates excellent agreement with the experimental data, validating both the numerical approach and the idealised theoretical model in the appropriate regime.

We can see that in both cases, there are peaks in the transmission spectra which correspond to the poles of the transmission matrix elements, and the sharp dip seen in the direct transmission is the antiresonance. 
We note that in the analytic model detailed earlier the system-bath couplings $\kappa$ and dampings $\gamma$ are assumed to be frequency independent. The resulting peaks in the transmission spectra will then be symmetric. 
We attribute the asymmetry of the peaks seen in Fig.~\ref{fig:Eres} to frequency dependent couplings and dampings captured in the COMSOL simulation but neglected in the simple coupled mode theory. 

\begin{figure*}[!htp]
\centering
\includegraphics[width=0.9\textwidth]{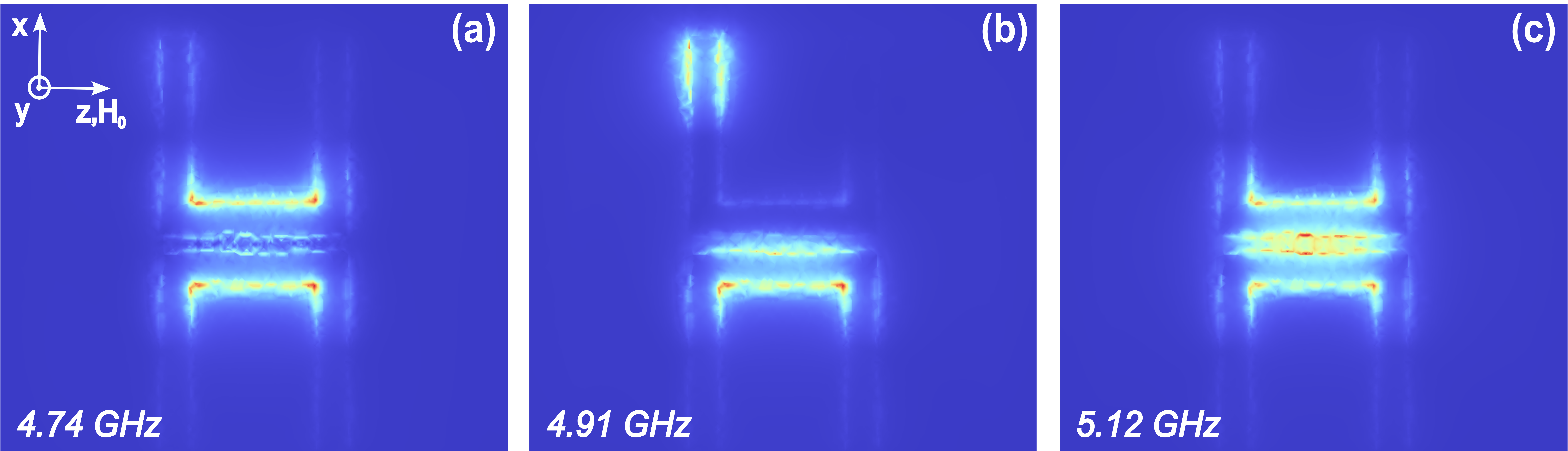}
\caption{Magnetic field intensity profiles for (a) the in-phase dual cavity mode (b) the antiresonance (c) the out-of-phase mode. The in-phase resonators have a magnetic field minima at the midpoint as discussed in the main text while for the out-of-phase resonators the field is at a maximum.}
\label{fig:fields}
\end{figure*}

In Fig.~\ref{fig:fields} we show the intensity profiles of the magnetic field for different frequencies of the excitation driving field (also obtained using COMSOL). 
In part (a), the magnetic field intensity profile is shown when the system is driven at $\omega_1/(2\pi)=4.74$ GHz. 
At this frequency, the resonators respond in phase and the resulting fields interfere destructively in between the resonators. 
We then expect that if the YIG sphere is centred between the resonators, the magnons do not couple to the low-frequency cavity photons ($g_1=0$). 
On the other hand, at $\omega_2/(2\pi)=5.12$ GHz [Fig.~\ref{fig:fields}(c)], the resonators respond out of phase, and the resulting fields interfere constructively. In this case, we expect the upper cavity photons couple strongly to the magnons. 
Additionally, at the antiresonance frequency, $\omega_a/(2\pi)=4.91$~GHz [Fig.~\ref{fig:fields}(b)], direct transmission is weak and indirect transmission is somewhat stronger $|t_i/t_d| = 2J/\gamma$; and hence, the field is strongest near the lower resonator. 
The antiresonance occurs because of destructive interference between the drive and because of feedback from the lower resonator which suppresses oscillations in the upper resonator creating a dark port at its output. The associated antiresonance becomes highly tunable when another drive is introduced and CPE of particular transmission pathways can be realised through control of relative phase and amplitude.

\subsection{Coherent Perfect Absorption and Extinction}
\label{sec:CPAandE}

As discussed, there are different scattering phenomena under the umbrella of CPE. 
In our resonator system, there are separate configurations of the drive fields that can be considered. 
Having now characterised the device, let us briefly pause to consider the conditions under which we can engineer some of the phenomena associated with the control of energy flow through the transmission pathways.

The empty resonators themselves are a lossy two-mode system, governed by a non-Hermitian Hamiltonian which may be engineered to achieve exceptional points, coherent perfect absorption, and reflectionless scattering modes  \cite{chongCPA, SweeneyRSM}. Often in photonic systems these phenomena have been realised through careful control of loss and gain. 

These phenomena are possible in the resonator even in the absence of the YIG sphere. In the proof-of-principle experimental realisation of the system, the intrinsic dampings of the modes exceed the possible gain from the drive fields, and these phenomena are not possible. However, we review them here to demonstrate potential applications of the system under consideration. The relevant input-output calculations are presented in Supporting Information~\ref{ap:IOempty}.

For clarity, we consider the empty resonators. We note these phenomena are possible with the YIG sphere present; however, additional tuning of the system parameters is required to achieve the necessary balanced loss and gain. One may obtain exceptional points in the empty resonators; however, this requires tuning of the system paramaters. As will be discussed in Sec. \ref{sec:TTD} and \ref{sec:EAM}, with multiple drive fields and the YIG sphere present, one may obtain exceptional points for any given system parameters.

From the input equation, the non-Hermitian Hamiltonian governing the system dynamics is 
\begin{align}
\label{eq:Hres}
\frac{\mathcal{H}_{res}}{\hbar} = \left[ \begin{array}{cc}
a_u^{\dag} & a_l^{\dag} \end{array} \right] 
\left[ \begin{array}{cc}
z_u & J  \\ 
J  & z_l  \\
\end{array} \right]
\left[ \begin{array}{c}
a_u \\ a_l \end{array} \right],
\end{align}
with $z_{u,l}=\omega_a - i\gamma_{u,l}/2$. The damping parameters are given in Eq. (\ref{eq:dampings}) of Supporting Information~\ref{ap:IOempty}. They include damping from the resonator ports as well as intrinsic damping of the resonator photons. This is a paradigmatic example of a photonic system in which exceptional points can occur \cite{Miri}. 

To begin, we consider reflectionless scattering modes. If resonator ports one and two are driven, the non-Hermitian Hamiltonian governing the dynamics of the driven-dissipative system is
\begin{align}
\frac{\mathcal{H}_{rsm}}{\hbar} = \left[ \begin{array}{cc}
a_u^{\dag} & a_l^{\dag} \end{array} \right] 
\left[ \begin{array}{cc}
\widetilde{z}_u & J  \\ 
J  & z_l  \\
\end{array} \right]
\left[ \begin{array}{c}
a_u \\ a_l \end{array} \right],
\end{align}
where $\widetilde{z}_u = \omega_a +i\widetilde{\gamma}_u/2$ and $\widetilde{\gamma}_u = \gamma_{u,\kappa} - \gamma_{u,int}$. The upper resonator now exhibits gain from the drive field, $\gamma_{u,\kappa}=2\pi(\kappa_{u1}^2+\kappa_{u2}^2)$, rather than the loss from the resonator ports seen in the input equation.

To achieve reflectionless scattering modes, there must be balanced gain and loss in the system. This requires $\gamma_{u,\kappa} = \gamma_{l,\kappa} + \gamma_{l,int} + \gamma_{u,int}$. One may achieve this provided the upper resonator couplings are stronger than the lower resonator couplings. Setting $\widetilde{\gamma}_u=\gamma_l=\gamma$, the frequencies at which reflectionless scattering modes occur are found to be
\begin{align}
\omega_{RSM}^{\pm} = \omega_a \pm \frac{1}{2} \sqrt{4 J^2-\gamma^2}.
\end{align}
One has reflectionless scattering modes provided $2J>\gamma$. An exceptional point occurs when $2J=\gamma$, and for $2J<\gamma$ the system transitions to a regime in which reflectionless scattering is no longer possible. 

Formally, the photonic system under consideration is similar to the cavity-magnonics system studied in Ref. \cite{zhang2017observation}, in which driven cavity photons are coupled to YIG magnons. Zhang \textit{et al.} achieved coherent perfect absorption, and by repositioning the YIG sphere to tune the magnon-photon coupling, they were able to drive the system through an exceptional point, and into a phase in which coherent perfect absorption is not possible. 
We note that, with excitation fields applied to Ports 1 and 2, the lower resonator photons play the role of the magnons in Ref. \cite{zhang2017observation} (as shown in in Eq.~(\ref{eq:Hsys2}) of Supporting Information C). 
As the lower resonator includes output channels, one obtains reflectionless scattering modes rather than coherent perfect absorption.

When all four resonator ports are driven, the non-Hermitian Hamiltonian governing the dynamics of the driven-dissipative system is
\begin{align}
\frac{\mathcal{H}_{ar}}{\hbar} = \left[ \begin{array}{cc}
a_u^{\dag} & a_l^{\dag} \end{array} \right] 
\left[ \begin{array}{cc}
\widetilde{z}_u & J  \\ 
J  & \widetilde{z}_l  \\
\end{array} \right]
\left[ \begin{array}{c}
a_u \\ a_l \end{array} \right],
\end{align}
where $\widetilde{z}_u$ is as before, and $\widetilde{z}_l=\omega_a+i\widetilde{\gamma}_l/2$ with $\widetilde{\gamma}_l = \gamma_{l,\kappa} - \gamma_{l,int}$. The system now exhibits gain from all four resonator ports. 

Defining $\widetilde{z}_{u,l} = \omega_a + i \widetilde{\gamma}_{u,l}/2$, the eigenvalues are
\begin{align}
z_{ar}^{\pm} = \frac{\widetilde{z}_u+\widetilde{z}_l}{2} \pm
\frac{1}{2} \sqrt{(\widetilde{z}_u-\widetilde{z}_l)^2+4J^2}.
\end{align}
If $\widetilde{\gamma}_u = -\widetilde{\gamma}_l$ there is balanced loss and gain, and the system can exhibit coherent perfect absorption at frequencies
\begin{align}
\omega_{CPA}^{\pm} =  \omega_a 
&\pm \frac{1}{2}\sqrt{4J^2-\biggr(\frac{\widetilde{\gamma}_u - \widetilde{\gamma}_l}{2}\biggr)^2}. 
\end{align}
If $\widetilde{\gamma}_u=\widetilde{\gamma}_l=0$, there will be coherent perfect absorption; however, this is not the most interesting scenario. In this case, the coherent perfect absorption frequencies are $\omega_{CPA}^{\pm} = \omega_a \pm J$ and there are no exceptional points. The more interesting scenario occurs if $\widetilde{\gamma}_u = -\widetilde{\gamma}_l \equiv \widetilde{\gamma} \neq 0$, allowing for the possibility of exceptional points. Note that in the absence of a drive field, the dynamics of the system is determined by $\mathcal{H}_{res}$. Driving the system prepares it in a particular state; if this state is an eigenstate of $\mathcal{H}_{ar}$ one has coherent perfect absorption and the system has $\mathcal{PT}$-symmetry \cite{Bender, Ozdemir, ZhangPT}. This occurs if $2J>\widetilde{\gamma}$. By separating the resonators (reducing $J$), the system is driven through an exceptional point below which ($2J<\widetilde{\gamma}$) the $\mathcal{PT}$-symmetry of the system is broken and coherent perfect absorption is no longer possible.

\section{Multi-Drive Systems}
\label{sec:TTD}

We now consider the four port system loaded with a YIG sphere placed between the two cavities. In Ref. \cite{grigoryan}, a scheme was proposed in which coupled cavity photons and magnons are independently driven, with the drive fields being attenuated and phase shifted relative to each other. This was experimentally realised in Refs. \cite{Boventer1, Boventer2}, and tunable level attraction and repulsion was observed in the reflection spectrum. Their system supports both modes, and antimodes that result from destructive interference in the transmission/reflection spectra \cite{RaoAntires, Bourcin, gardin2025level}. 
The two excitation driving fields allow the antimodes to be tuned between a regime in which there is level repulsion, and one in which there is level attraction.

Our system differs from Refs. \cite{grigoryan, Boventer1, Boventer2, gardin2025level} in that we have two cavity modes, four resonator ports, and the magnon mode is not driven directly. 
Phase shifting and attenuating the signals sent to the driven resonators allows tunability between antimode level repulsion and attraction, exceptional points \cite{Heiss}, and coherent perfect extinction of transmitted or reflected light \cite{guo2023singular}. 

The experimental parameters relevant to the system under consideration are listed in Table \ref{tab:expt}. We base this on the single drive case with the YIG sphere centred between the resonators where $g_{1} \sim 0$ (see Supplemental Information~\ref{ap:estimation} for details on the parameter estimation). We note that there is some change in the resonator properties owing to the effect of the small ceramic rod upon which the $0.2$ mm YIG sphere is mounted. When the field intensity is prominent between the resonators, the dielectric contribution from the ceramic rod lowers the resonant frequency of the mode $\omega_{2}$ \cite{Pozar}, also reducing the effective $J$ value. 
The high frequency magnon-polariton modes satisfy $\gamma_m < g_2 < \gamma_{int}$, which means the system is operating in the magnetically induced transparency regime \cite{Zhang}.

\begin{table}[!htp]
\centering
\begin{tabular}{|c|c|}
\hline
$\omega_a/(2\pi)$           & $4.81$ GHz             \\ \hline
$J/(2\pi)$                  & $160.91$ MHz           \\ \hline
$\gamma_{\kappa}/(2\pi)$    & $2.08$ MHz             \\ \hline
$\gamma_{int}/(2\pi)$       & $61.37$ MHz            \\ \hline 
$\gamma_m/(2\pi)$           & $11.65$  MHz           \\ \hline
$g_{2}/(2\pi)$              & $18.14$ MHz            \\ \hline
\end{tabular}
\caption{Experimental parameters when the YIG sphere is centred between the resonators. The inductively coupled ($J$) open-ring resonators operate at $\omega_a$. The total damping from the resonator ports is $\gamma_{\kappa}=4\pi \kappa^2$ with $\kappa = 1.02 \sqrt{\text{MHz}}$. The low and high frequency cavity modes, $\omega_{1,2} = \omega_a \mp J$, have equal intrinsic damping of $\gamma_{int}$, and $\gamma_m$ is the damping of the YIG sphere. }
\label{tab:expt}
\end{table}

Consider our multi-drive system in which the input is to ports one and three, with $x_1^{IN}=\delta e^{i\Phi} x_3^{IN}$. From the S-matrix [Eq. (\ref{eq:Smatrix})], the output at port two is determined by
\begin{align}
\label{eq:s2}
S_2 = t_{du} - \delta e^{i\Phi} t_i.
\end{align}
The poles of this function are the modes of the system, and its zeros are the scattering channel dependent antimodes. As noted in Sec. \ref{sec:Smatrix}, one may obtain coherent perfect extinction of the output by setting $\delta e^{i\Phi} = t_{du}/t_i$. Similar reasoning applies to the other output ports. With multiple transmission pathways, the multiple drives provide a means of using destructive interference to extinguish output from any given resonator port. As this is possible at any frequency (not just a coherent perfect absorption frequency), the device is suitable for interferometry at any frequency.

\begin{figure*}[!htp]
\centering
\includegraphics[width=\textwidth]{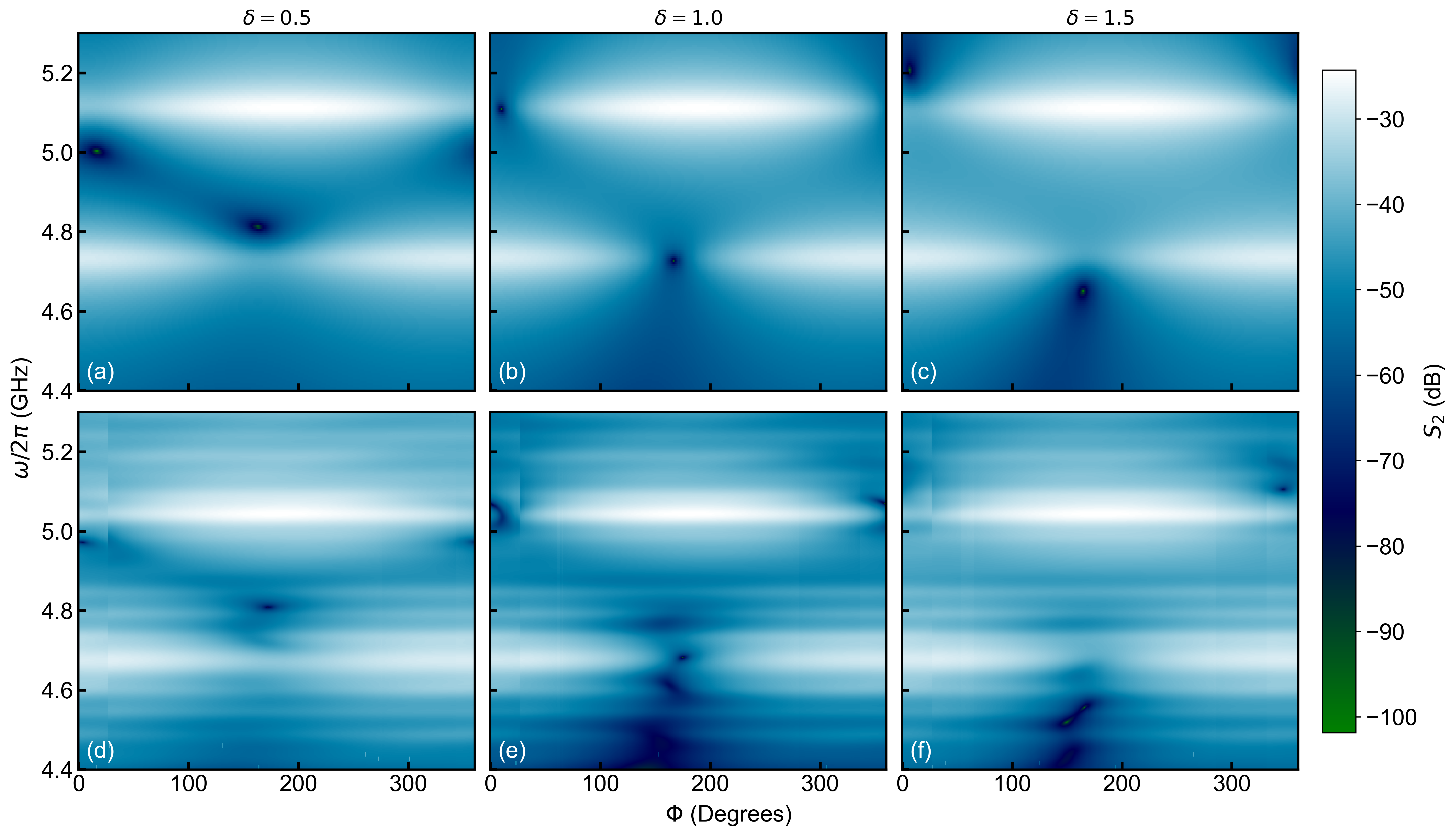}
\caption{COMSOL simulation and experimental data showing the effect of tuning the relative phase and amplitude of inputs to ports one and three with the output measured at port two. The experimental data, particularly for $\delta =$ 1.0 and 1.5, contains additional spurious antiresonances occuring due to interference with the external circuitry that is unaccounted for in the simulations.}
\label{fig:phaseantituning}
\end{figure*}

We demonstrate an example of this destructive interference tuning in Fig.~\ref{fig:phaseantituning}, which shows the empty resonator's transmission through port two ($S_2$) as the relative phase and amplitude into ports one and three is tuned. In the experiment, despite the presence of spurious antiresonances resulting from interference with the measurement setup, we can tune the system to the points corresponding to the simulations. This allows us to calibrate the phase determined by the simulation with that of the fabricated resonator system, and investigate the effect of magnon coupling on the antimodes (see Supplemental Information~\ref{ap:calibration}).

Using the coupled-mode model, we explore the behaviour expected with the YIG sphere present using input-output theory. Defining  
\begin{align}
\label{eq:zbar1}
\overline{z}_1&=\frac{z_1(1-\delta e^{i\Phi})+z_2(1+\delta e^{i\Phi})}{2},
\end{align}
with $z_{1,2} = \omega_{1,2} - i\gamma_{1,2}/2$, and $\overline{z}_2=z_2 = \omega_m - i\gamma_m/2$, one finds the complex antimode frequencies to be
\begin{align}
\label{eq:zpmbar}
&\overline{z}_{\pm} = \frac{\overline{z}_1+\overline{z}_2}{2} \pm \frac{1}{2}
\sqrt{(\overline{z}_1-\overline{z}_2)^2+4\overline{g}^2},
\\ \nonumber &\text{where}\qquad
\overline{g}^2 = g_l (g_l+\delta e^{i\Phi}g_u).
\end{align}
The antimodes $\overline{z}_{\pm}$ may be viewed as the result of a pair of coupled antimodes $\overline{z}_{1,2}$, with complex coupling constant $\overline{g}$. 

\begin{figure*}[!ht]
\centering
\includegraphics[width=\textwidth]{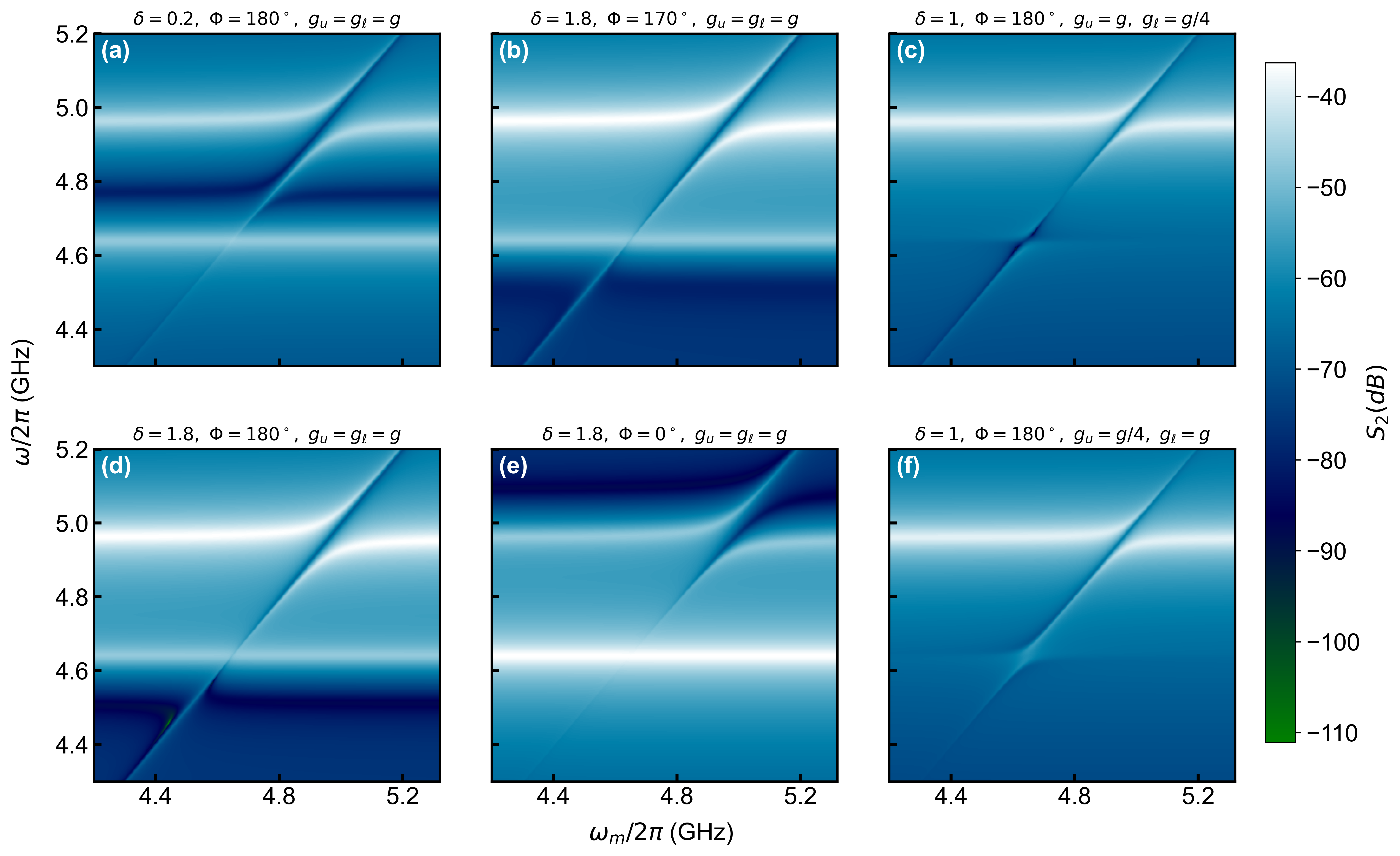}  
    \caption{Transmission spectra showing the effect of tuning the relative amplitude of the drives $\delta$, the relative phase of the drives $\Phi$, and tuning the coupling strengths, $g_l$ and $g_u$. We assume a pair of $\omega_a/(2\pi)=4.8$ GHz resonators inductively coupled with strength $J/(2\pi)=160$ MHz. The coupling strengths are measured in units of $g/(2\pi)=50$ MHz. Parts (a) and (d) show the effect of increasing $\delta$ with $\Phi=180^{\circ}$. The system exhibits antimode level repulsion for $\delta<1$, and level attraction for $\delta>1$. Parts (b) and (e) show the affect of tuning $\Phi$. Going from $0^{\circ} \ \rightarrow \ 180^{\circ}$ can tune the system between level attraction and repulsion with the linewidth of the horizontal antimode becoming large at intermediate values of $\Phi$. Parts (c) and (f) show the affect of shifting the YIG sphere with $\delta=1$ and $\Phi=180^{\circ}$. The horizontal antimode is then degenerate with the lower cavity mode. With the YIG sphere closer to the lower resonator the antimodes exhibit level repulsion. Shifting the YIG sphere towards the upper resonator tunes the system into a level attraction regime. See the text for more details.}
    \label{fig:LALR}
\end{figure*}

One may tune between level attraction and repulsion by tuning $\delta$ or $\Phi$, or by tuning the coupling constants $g_u$ and $g_l$. Tuning of the coupling constants is achieved by shifting the position of the YIG sphere. Note that it is the presence of the YIG sphere that leads to the complex coupling between the antimodes. This may be utilised to achieve antimode level attraction. The transmission spectra in Fig. \ref{fig:LALR} show the tunability of the antimodes. We assume $\omega_a/(2\pi) = 4.8$ GHz resonators, with inductive coupling $J/(2\pi)=160$ MHz. The system bath couplings are set to $\kappa=1 \ \sqrt{MHz}$. To better show the level attraction and repulsion we set $g/(2\pi)=50$ MHz, which is larger than the experimental value, and to sharpen the spectral features we set the intrinsic damping to $\gamma_{int}/(2\pi)=30$ MHz, which is about half the experimental value. The damping of the magnon mode is set to $\gamma_m/(2\pi)=10$ MHz.

In parts (a) and (d) of Fig.~\ref{fig:LALR} we show the effect of tuning the relative amplitude of the drive field $\delta$ with the YIG sphere centred between the resonators ($g_u=g_l=g$). 
The phase is set to $\Phi=180^{\circ}$ meaning the inputs from the vector network analyzer arrive out of phase; the antimode coupling constant is then $\overline{g}^2=g^2(1-\delta)$. 
With $\delta<1$, the antimodes exhibit level repulsion in the region between the cavity modes. 
With $\delta>1$ one finds $\overline{g}^2<0$, and the modes exhibit level attraction below the low-frequency cavity mode. At Port 4, the system exhibits converse behaviour, viz., there is antimode level attraction when $\delta<1$ and repulsion when $\delta>1$. 
We note that although the feature seen in the spectrum shown in Fig.~\ref{fig:LALR}(d) is characteristic of level attraction, with our chosen system parameters there are no exceptional points, and the antimodes are only degenerate at a single point. 
The linewidth of the antimodes rapidly increases in the level attraction region, so these features are not apparent in the spectrum. 
We will discuss exceptional points in Sec. \ref{sec:EAM}.

The effect of tuning $\Phi$ is shown in Figs.~\ref{fig:LALR}(b) and (e). 
As in parts (a) and (d), the YIG sphere is centred between the resonators; the antimode coupling is then $\overline{g}^2 = g^2(1+\delta e^{i\Phi})$. With $\delta=1.8$, as one reduces $\Phi$ from $180^{\circ}$ the antimodes broaden considerably and the level attraction region shrinks as seen in part (b). The horizontal antimode shifts up in frequency reaching $4.8$ GHz at $\sim 90^{\circ}$. 
As $\Phi$ is reduced below $90^{\circ}$ the horizontal antimode continues to shift up in frequency, its linewidth decreases, and eventually level repulsion develops as shown in part (e). 

Finally, in Figs.~\ref{fig:LALR} (c) and (f) we tune the coupling parameters with $\delta=1$ and $\Phi=180^{\circ}$. As $g_u \neq g_l$ both the low and high frequency cavity modes hybridise with the magnons. With $\delta=1$ the low-frequency cavity mode and the horizontal antimode are degenerate. Tuning the couplings does not change the frequencies of the antimodes; however, it changes the width of the level attraction/repulsion regions. 
In part (c) the YIG sphere is located near the upper resonator ($g_u>g_l$), and the antimodes exhibit level repulsion in the region between the avoided level crossing in the mode spectrum. If the YIG sphere is shifted towards the lower resonator the level attraction region shrinks, and with ($g_l>g_u$) both the modes and the antimodes repel one another.


\begin{figure*}[!htb]
\includegraphics[width=\textwidth]{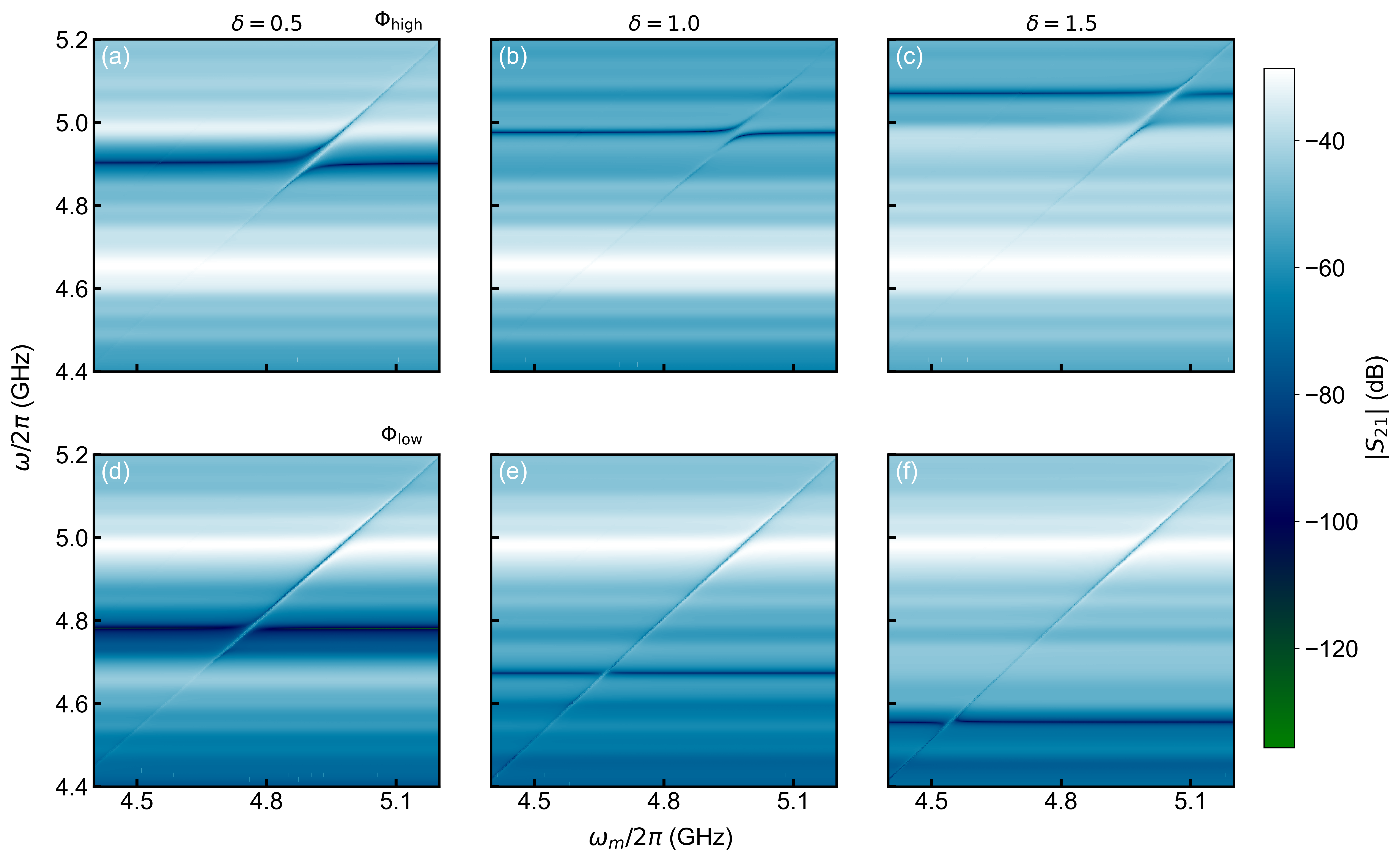}
\caption{Transmission through port two with ports one and three driven with relative amplitude $\delta$ and phase $\Phi$. We see the effects of phase and amplitude tuning on the antiresonance whilst the sphere is positioned in the centre of the two resonators. Left to right, parts (d), (e), (f) show the effect of amplitude tuning when $\Phi \approx 180$. This corresponds to parts (a) and (d) of Fig. \ref{fig:LALR}. Part (a), (b), (c) show the effect of amplitude tuning on the avoided level crossing when $\Phi \approx 0$. Top to bottom shows the effect of phase tuning. Parts (f) and (c) correspond to (d) and (e) of Fig. \ref{fig:LALR}. See the text for further details.}
\label{fig:spheremid}
\end{figure*}

In Fig. \ref{fig:spheremid}, we present the equivalent experimental results of effect of relative phase and amplitude tuning on the antimodes obtained using the experimental set up given in Fig.~\ref{fig:combined1}(a). Level attraction is obtained for values of $\Phi$ near 180$^\circ$ (see Supplemental Information~\ref{ap:calibration}) with $\delta>1$ [part (f)]. 
As $\delta$ is reduced [parts (f), (e), and (d)] the system transitions to a regime in which there is level repulsion ($\delta<1$). 
This corresponds to parts (d) and (a) of Fig. \ref{fig:LALR}. With $\delta = 1$, as seen in both parts (b) and (e), the antimode suppresses both the low-frequency uncoupled cavity mode and the high-frequency magnon-polariton modes. The transition between repulsion and attraction takes place as the antimode crosses the low-frequency cavity mode. Parts (a), (b), and (c) show the effect of amplitude tuning for values of $\Phi$ near $0^{\circ}$. We see the antiresonance is driven across the upper cavity mode; however, there is no transition between level attraction and repulsion.

Top to bottom Fig. \ref{fig:spheremid} shows the effects of phase tuning at fixed amplitude. Parts (f) and (c) correspond to parts (b) and (e) of Fig. \ref{fig:LALR}. Reducing the frequency from $\Phi \approx 180^{\circ}$ to $\Phi \approx 0^{\circ}$ shifts the antiresonance up in frequency, with significant broadening occurring at intermediate values of $\Phi$. As is apparent in the figure, with $\delta<1$ the antiresonance is located between the cavity modes. With $\delta>1$, the antimodes can occur outside the frequency range bounded by the cavity modes, and they may exhibit level attraction.

Having discussed level attraction and repulsion in the antimode spectrum, we now turn to exceptional antimodes. As noted previously, there are no exceptional points associated with the level attraction region seen in Figs. \ref{fig:LALR} and \ref{fig:spheremid}. However, one may always use $\delta$ and $\Phi$ to tune the system to an exceptional point, as we show in Sec. \ref{sec:EAM}. 

\section{Exceptional Antimodes}
\label{sec:EAM}

The exceptional points we consider here are unconventional in the sense that they do not correspond to poles or zeros of the full $S$-matrix. Rather, we consider exceptional points in the antimode spectrum at a single output port. Consider, for example, output from port two in the scheme discussed in Sec. \ref{sec:TTD}. One may write Eq. (\ref{eq:s2}) as
\begin{align}
S_2 = -i2\pi \kappa^2 \frac{\det \left[\omega \mathbbm{1} -h_2\right]}
{\det \left[\omega \mathbbm{1} - h_{res}\right]}
\end{align}
where $\Omega= \omega \mathbbm{1} - h_{res}$ is a matrix that determines the resonances of the system, and 
\begin{align}
h_2 = \left[ \begin{array}{cc}
\overline{z}_1 & \overline{g}  \\ 
\overline{g}  & \overline{z}_2  \\
\end{array} \right].
\end{align}
This matrix determines the antiresonance frequencies of the multi-drive system at port two;
its eigenvalues are $\overline{z}_{\pm}$ [Eq. (\ref{eq:zpmbar})], and the right eigenvectors are 
\begin{align}
r_{\pm} = \frac{1}{c_{\pm}}
\left[ \begin{array}{c}
\overline{z}_{\pm} - \overline{z}_2   \\ \overline{g}   \\
\end{array} \right] 
\end{align} 
where
\begin{align}
c_{\pm}^2 = [\overline{z}_{\pm}^* - \overline{z}_2^*,\ \overline{g}^*] \left[ \begin{array}{c}
\overline{z}_{\pm} - \overline{z}_2   \\ \overline{g}   \\
\end{array} \right].
\end{align}
When $\overline{z}_+=\overline{z}_-$ the two eigenvectors coalesce at an exceptional point.

From Eq. (\ref{eq:zpmbar}), the exceptional point condition is
\begin{align}
(\overline{z}_1-\overline{z}_2)^2+4\overline{g}^2 = 0
\end{align}
Using Eq. (\ref{eq:zbar1}), one may obtain a quadratic polynomial, $p(x)=ax^2+bx+c=0$, with $x=\delta e^{i\Phi}$ and
\begin{align}
a&=(z_1-z_2)^2 \\ \nonumber
b&=-2(z_1+z_2-2z_m)(z_1-z_2)+16g_ug_l \\ \nonumber
c&=(z_1+z_2-2z_m)^2+16g_l^2.
\end{align}
Thus, for any given system parameters, one may use $\delta$ and $\Phi$ to tune the system to an antimode exceptional point. 

\begin{figure}[!htb]
	\centering
	\mbox{\includegraphics[width=7cm]{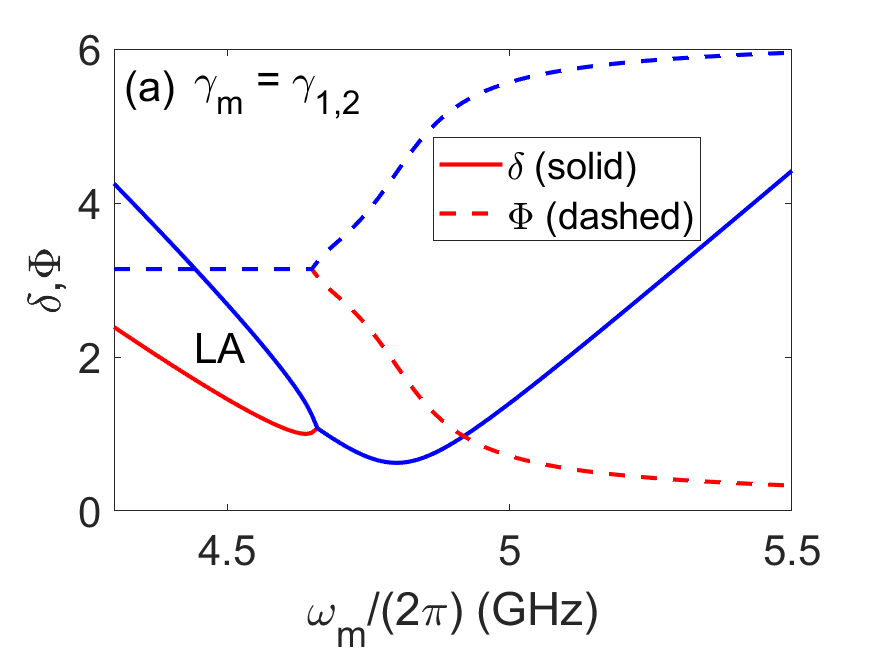}}      
    \mbox{\includegraphics[width=7cm]{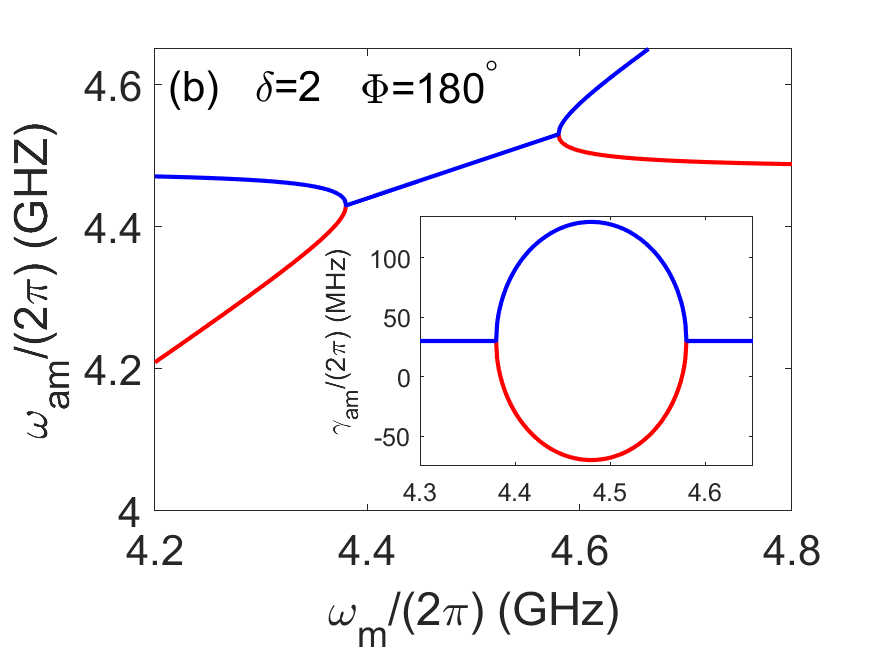}}   
    \hspace{10cm}
    \mbox{\includegraphics[width=7cm]{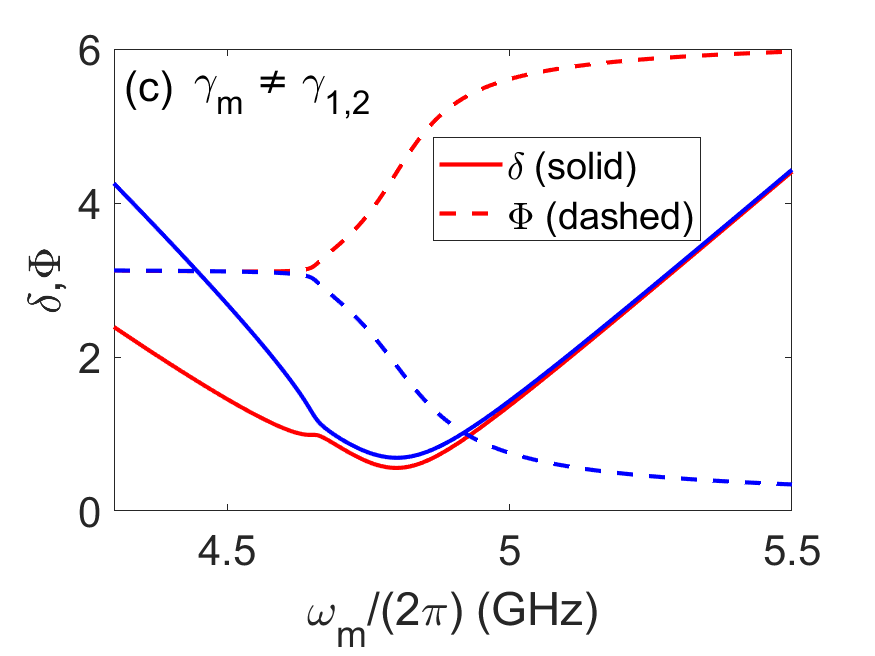}}
    \mbox{\includegraphics[width=7cm]{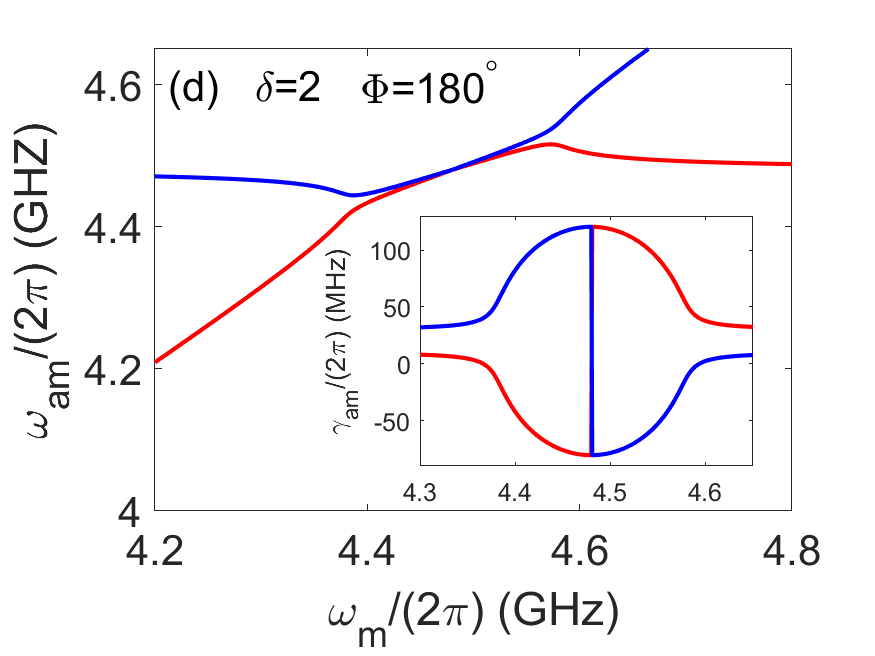}}
    \caption{Antimode exceptional point map for output at port two ($S_2=t_{du}-\delta e^{i\Phi}t_i$) determined by the polynomial $p(x)$ (see text). The YIG sphere is centred between the resonators ($g_u=g_l=2\pi \times 50$ MHz) In part (a) we set $\gamma_m = \gamma_{1,2}$, in which case the polynomial determining $\delta e^{i\Phi}$ has two real roots ($\Phi=180^{\circ}$ in the level attraction region), or the roots come in complex conjugate pairs. Level attraction occurs in the region bounded by the solid lines which determine $\delta_{EP}^{\pm}$ when $\Phi_{EP}=180^{\circ}$. Part (b) shows antimode frequencies $\omega_{am}$, and linewidths $\gamma_{am}$, when $\delta=2$ and $\Phi=180^{\circ}$. In part (c) we set $\gamma_m = 2\pi \times 10\ \text{MHz} \neq \gamma_{1,2}$. One may still obtain exceptional points for any given system parameters; however, in the ``level attraction" region one has two distinct values of $\Phi_{EP} \neq 180^{\circ}$. As a function of $\omega_m$, with $\Phi$ fixed, the antimodes are no longer degenerate in the ``level attraction" region, although they do exhibit significant broadening as shown in part (d).}
    \label{fig:EPmap}
\end{figure}

In Fig. \ref{fig:EPmap} (a) we show an exceptional point map for a resonator with the same parameters as in Fig. \ref{fig:LALR}, apart from $\gamma_m$ which we set equal to $\gamma_{1,2}$; the polynomial $p(x)$ then has real coefficients. The YIG sphere is centred between the resonators ($g_u=g_l=2\pi \times 50$ MHz). Below $\omega_m^c/(2\pi)=4.66$ GHz, one finds $p(x)$ has two real roots with $\Phi$ fixed at $180^{\circ}$, and above $\omega_m^c$ the roots are complex conjugate pairs. The critical point $\omega_m^c$ is itself an exceptional point associated with the companion matrix of $p(x)$, as will be discussed further below. The region bounded by the solid lines [$\delta_{EP}(\Phi=180^{\circ})$], marked LA in the figure, is the level attraction region.

In Fig. \ref{fig:EPmap} (b) we show the antimode frequencies and linewidths ($\omega_{am}$ and $\gamma_{am}$) with $\delta=2$ and $\Phi=180^{\circ}$. The level attraction region is bounded by a pair of exceptional points that occur at complex frequencies. Generally, the antimode frequencies are complex, with significant broadening of the modes occurring in the level attraction region. Coherent perfect extinction occurs at two points in the level attraction region, distinct from the exceptional points, where the linewidth of an antimode crosses zero.

More generally, if $\gamma_{1,2} \neq \gamma_m$, the roots of $p(x)$ are complex, and the phase at which the exceptional points occur ($\Phi \neq 180^{\circ}$) will vary with $\omega_m$. In Fig. \ref{fig:EPmap} (c) we show an exceptional point map for the system parameters used in Fig. \ref{fig:LALR} ($\gamma_m=2\pi \times 10 \ \text{MHz} \neq \gamma_{1,2}$). For fixed $\delta$ and $\Phi$ the antimodes are no longer degenerate in the ``level attraction" region. With, for example, $\delta=2$ and $\Phi=180^{\circ}>\Phi_{EP}$ there is still a region in which nearly degenerate antimodes exhibit significant broadening [see Fig. \ref{fig:EPmap} (d)], but the antimodes cross at a single point, and there are no exceptional points. The ``level attraction" shown in Fig. \ref{fig:LALR} is of this form. To obtain an exceptional point at either end of the ``level attraction" region one must tune $\Phi$.

\begin{figure}[!htb]
	\centering
	\mbox{\includegraphics[width=6cm]{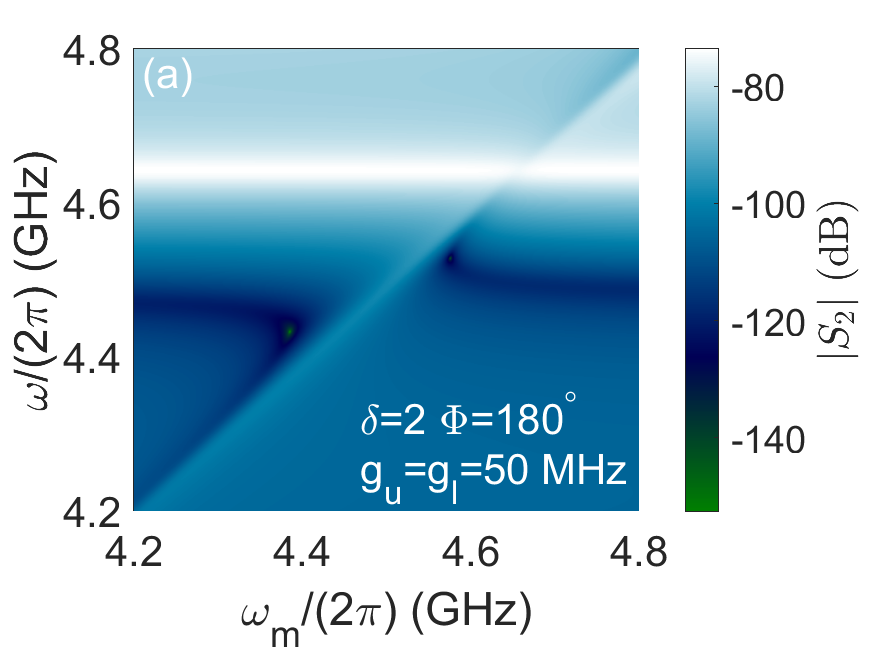}}    
    \mbox{\includegraphics[width=6cm]{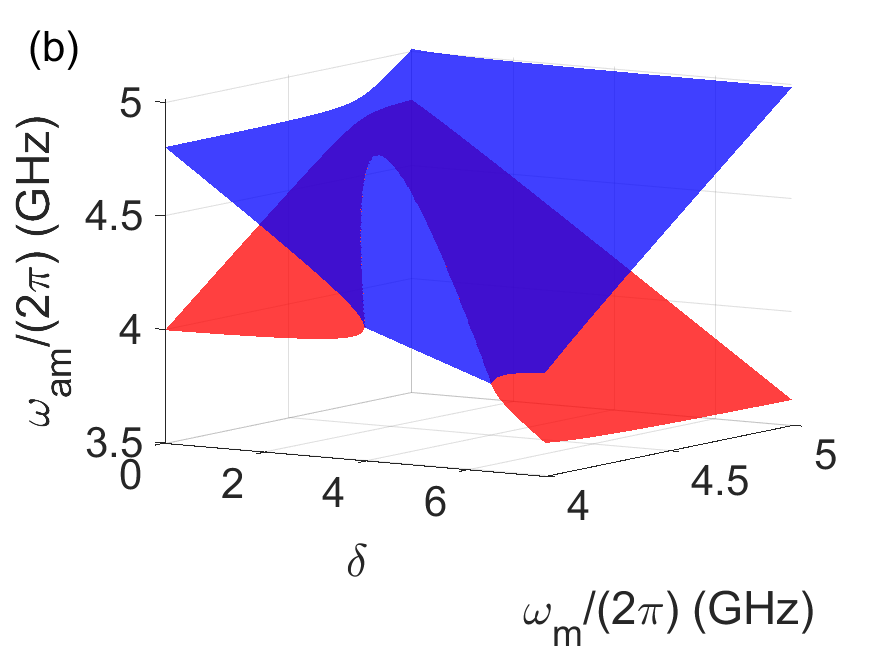}}   
    \mbox{\includegraphics[width=6cm]{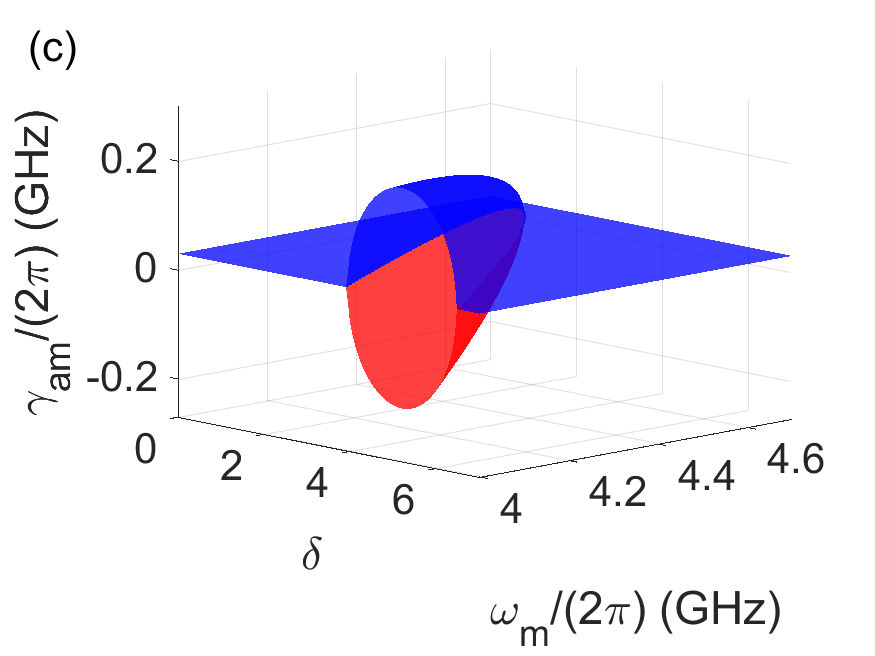}} 
    \caption{Calculated output spectrum [part (a)] at resonator port two ($S_2=t_{du}-\delta e^{i\Phi} t_i$) and antimode Riemann sheets. The parameters used are the same as in Fig. \ref{fig:EPmap} (b). With $\gamma_m = \gamma_{1,2}$ the modes are degenerate in the level attraction region. In the experiment, $\gamma_m \neq \gamma_{1,2}$, and one must tune $\delta$ and $\Phi$ to obtain exceptional points. In part (b) we show the antimode frequencies as a function of $\delta$ and $\omega_m$ with $\Phi=180^{\circ}$. The two surfaces meet in the level attraction region which is bounded by a line of exceptional points. In part (c) one sees the substantial broadening of the antimode linewidths in the level attraction region.}
    \label{fig:EP}
\end{figure}

The spectrum associated with the parameters used in Fig. \ref{fig:EPmap} (b) is shown in Fig. \ref{fig:EP}, along with antimode Riemann surfaces. The antimode is not visible in the level attraction region due to its increased linewidth. The Riemann surfaces shown are the antimode frequencies and linewidths with $\Phi=180^{\circ}$ as one varies $\delta$ and $\omega_m$. The level attraction region shown in Fig. \ref{fig:EP} (b) is bounded by a line of exceptional points. Outside the level attraction region the linewidths of the antimodes are degenerate and relatively small, $\gamma_{am}/(2\pi) \approx 30$ MHz. As one tunes the system to the center of the level attraction region the linewidth broadens rapidly, as shown in Fig. \ref{fig:EP} (c).

\begin{figure}[!htb]
	\centering
	\mbox{\includegraphics[width=7cm]{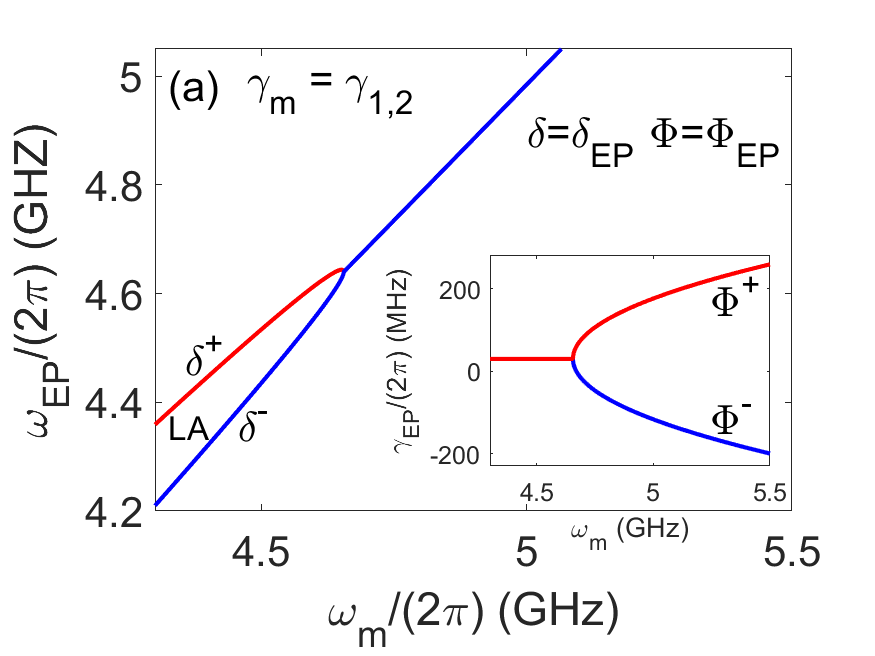}}      
    \mbox{\includegraphics[width=7cm]{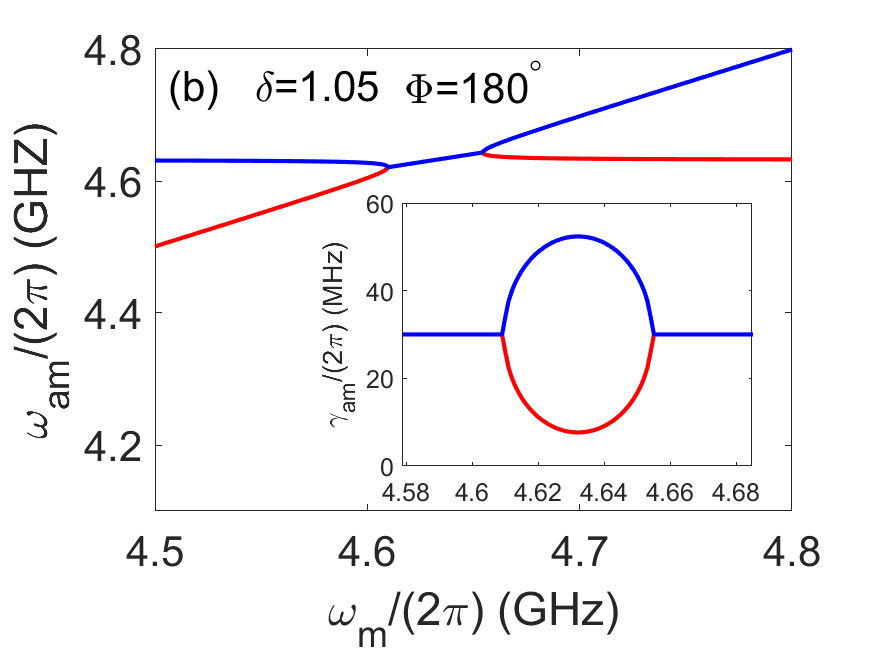}} 
    \hspace{10cm}
    \mbox{\includegraphics[width=7cm]{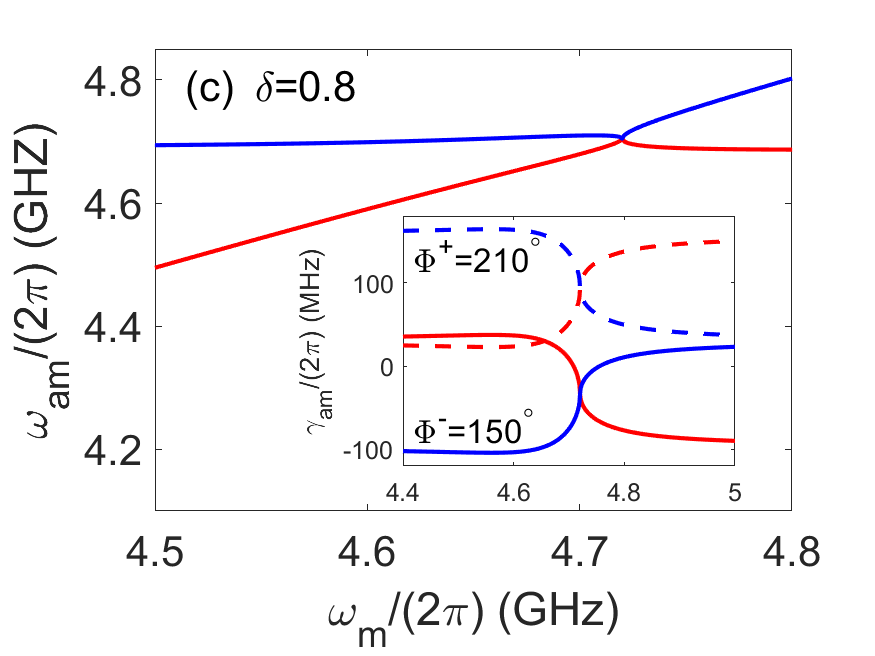}}
    \mbox{\includegraphics[width=7cm]{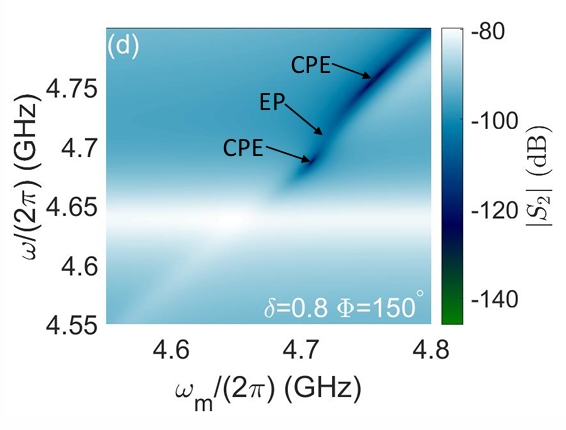}}
    \caption{Antimode frequencies, linewidths, and exceptional points using the same parameters as in Fig. \ref{fig:EPmap} (a). Ports one and three are driven, and the output is measured at port two. The YIG sphere is centred between the resonators ($g_u=g_l=2\pi \times 50$ MHz), and $\gamma_m=\gamma_{1,2}$. Part (a) shows the exceptional point frequencies as a function of $\omega_m$ with $\delta$ and $\Phi$ tuned to their exceptional point values. At $\omega_m^c/(2\pi)=4.66$ GHz the two exceptional antimodes coalesce at an exceptional point with $\delta=1$ and $\Phi=180^{\circ}$. As $\delta \rightarrow 1$ from above, the level attraction region shrinks, as shown in part (b). For $\delta<1$, a single exceptional point at $\omega_m>\omega_m^c$ occurs in the spectrum at two different relative phase differences $\Phi_{EP}^{\pm}$ as shown in part (c). The calculated spectrum at output port two with $\delta=0.8$ and $\Phi=\Phi_{EP}^- = 150^{\circ}$ is shown in part (d). The exceptional point doesn't show up as a clear spectral feature due to its increased linewidth; however, points at which there is coherent perfect extinction (the antimode has zero linewidth) show up clearly.}
    \label{fig:wEP}
\end{figure}

As previously noted, the bifurcation seen in the exceptional point map in Fig. \ref{fig:EPmap} (a) at $\omega_m^c/(2\pi)=4.66$ GHz is itself an exceptional point. If $\delta$ and $\Phi$ are tuned to their exceptional point values, then the antimode frequencies, $\omega_{EP}(\omega_m)|_{\delta_{EP},\Phi_{EP}}$ coalesce at $\omega_m^c$. The linewidth of $\omega_{EP}$ increases rapidly when $\omega_m>\omega_m^c$, as shown in Fig. \ref{fig:wEP} (a). Below $\omega_m^c$ the exceptional points occur at $\Phi_{EP}=180^{\circ}$ and $\delta_{EP}=\delta_{EP}^{\pm} > 1$; above $\omega_m^c$ the exceptional points occur at a fixed $\delta_{EP}$ and two different values of $\Phi_{EP}=\Phi_{EP}^{\pm}$. The linewidth of the exceptional points above $\omega_m^c$ show substantial broadening.

At fixed $\delta>1$ and $\Phi=180^{\circ}$, the two exceptional point frequencies $\omega_{EP}^{\pm}<\omega_m^c$ mark the boundaries of the level attraction region. As $\delta \rightarrow 1$, the level attraction region shrinks, as shown in Fig. \ref{fig:wEP} (b), and with $\delta<1$ one may obtain a single exceptional point at $\omega_{EP}>\omega_m^c$ and $\Phi=\Phi_{EP}^{\pm}$. The antimode frequencies are degenerate for the two solutions $\Phi_{EP}^{\pm}$; however, the linewidths are not. In Fig. \ref{fig:wEP} (c) we set $\delta=0.8$, in which case $\Phi_{EP}^-\approx 150^{\circ}$ and $\Phi_{EP}^+ \approx 210^{\circ}$. Note that the linewidth of the $\Phi_{EP}^-$ solution crosses zero whereas the $\Phi_{EP}^+$ solution does not. When the linewidth crosses zero there is coherent perfect extinction of the transmitted light which shows up clearly in the spectrum, as shown in Fig. \ref{fig:wEP} (d). The linewidth broadens at the exceptional point, and it does not show up as an easily discernible spectral feature. In theoretical calculations frequency cuts at fixed $\omega_m$ clearly show the features.

These novel, interference based, exceptional points will exhibit the same sensitivity to perturbations as the more conventional exceptional points that occur when poles or zeros of the full $S$-matrix coalesce. As one may use the two drive fields to achieve these exceptional points for any given system parameters, this may prove advantageous for sensing applications.

\newpage

\section{Discussion}

We have engineer a four-port, three-mode, cavity-magnonics system where interference based phenomena including exceptional points are released. 
We also build a robust theoretical framework which reveals physical insight into the system, namely, demonstrating that the modes of the system are associated with poles of the S-matrix and the antimodes are associated with its zeros. 
In this model, although we can write down a matrix describing the coupled antimodes, its eigenvectors are not eigenstates of the system. The antimodes are interference based; they depend on the drive fields and the scattering channel of interest. 
The antimodes do not couple to the modes, rather, they are the result of destructive interference between the modes. One may utilise multiple drive fields to obtain destructive interference at any given resonator port, a form of coherent perfect extinction. 
The tunability of the driving excitation vector fields allow this coherent perfect extinction to be achieved for any given system parameters, and allow the antimodes to be tuned between regimes in which there is level attraction and repulsion. Furthermore, for any given system parameters, multiple drive fields may be used to tune the system to antimode exceptional points.

Although we have not discussed topological aspects of our four port system, there are some interesting aspects worth noting. First, as noted in the introduction, scattering phenomena such as coherent perfect absorption and extinction are topologically nontrivial. The S-matrix is a function of its parameters. If, for example, one calculates the winding number of the S-matrix for a simple loop in its parameter space, the value of this topological invariant determines whether or not coherent perfect absorption may be achieved for parameters within the loop  \cite{guo2023singular}. 

Second, if a non-Hermitian system is prepared in a particular state, and the system parameters are tuned around a closed loop encircling an exceptional point, topological energy transfer (mode swapping) is possible \cite{Dembowski2001, xu2016topological, LambertModeSwitch, ZhangModeSwitch}. Here, as we are dealing with absorbing exceptional points, the relevant eigenvectors are not eigenstates of the system. The physical significance, if any, of tuning the system parameters around an absorbing exceptional point remains to be explored. 

Finally, as noted earlier, the topology of non-Hermitian systems allows for braiding of the complex eigenvalues \cite{Wojcik, Patil, Chavva}. In Ref. \cite{rao2024braiding}, a single port system was considered in which the cavity mode couples to a pair of magnon modes. These authors have shown that complex zeros and poles of the S-matrix may be braided with opposite chiralities. Braiding of eigenvalues in the system under consideration remains a subject for future work.

At present, in this particular case study, we have found the primary advantage of the system may be for sensing. If the system is tuned to an exceptional point, a perturbation of the form $\mathcal{H}=\mathcal{H}_{EP}+\epsilon \mathcal{H}'$ leads to a splitting of the eigenvalues $\propto \sqrt{\epsilon}$, whereas at a conventional degeneracy (a diablolic point) the splitting is $\propto \epsilon$. Thus, for small $\epsilon$, the system shows enhanced sensitivity to perturbations \cite{WiersigReview}. In Refs. \cite{ZhangEP3, Zhong}, systems having surfaces of exceptional points were presented. These exceptional surfaces combine enhanced sensitivity with robustness necessary for applications. The multi-drive system under consideration provides similar robustness as the amplitude and phase of the drive fields can be adjusted to compensate for fabrication errors, and the system can be tuned to an exceptional point for any given system parameters. This will allow for schemes similar to that proposed in Ref. \cite{mao2024exceptional} for sensitive phase detection of optical signals to be introduced to the microwave domain.

\section{Conclusions}

We have presented a theoretical model for, and the experimental realization of, a cavity-magnonics system showing remarkable versatility and utility. Multiple drive fields and transmission pathways allow for coherent perfect extinction of the output at any given resonator port. This can be achieved for any given system parameters, at any relevant frequency, making the system suitable for interferometry.

More significantly we have shown that, at a chosen resonator port, interference based antimodes exhibit tunable, complex couplings. This may be utilised to achieve absorbing exceptional points in the output spectrum. As the exceptional antimodes seen in the output spectrum share the same sensitivity to perturbations as resonant exceptional points, this may prove useful in future sensing applications.

\medskip
\textbf{Supporting Information} \par 
Supporting Information is available from the Wiley Online Library or from the author.

\medskip
\textbf{Acknowledgements} \par 

This work was supported by the University of Manitoba, the Natural Sciences and Engineering Research Council of Canada (NSERC RGPIN 05011-18, RGPIN-2018-05012, and ALLRP 597808 - 24), the Canadian Foundation for Innovation (CFI) John R. Evans Leaders Fund, the Engineering and Physical Sciences Research Council (EPSRC), and the Science and Technology Facilities Council (STFC).
M.A.S. acknowledges funding from EPSRC under grant number EP/S023321/1.

For the purpose of open access, the authors have applied a Creative Commons Attribution (CC BY) licence to any Author Accepted Manuscript version arising from this submission.

\medskip
\textbf{Data Availability Statement} \par

Data is available from the authors upon request. 

\medskip

%
\bibliographystyle{MSP}
\bibliography{antiresbib}

\begin{thebibliography}{100}
\providecommand{\url}[1]{\texttt{#1}}
\providecommand{\urlprefix}{URL }

\bibitem{FlebusReview}
B.~Flebus, et~al.,
\newblock \emph{J. Phys. Condens. Matter} \textbf{2024}, \emph{36} 1.

\bibitem{LQ}
D.~Lachance-Quirion, Y.~Tabuchi, S.~Ishino, A.~Noguchi, T.~Ishikawa, R.~Yamazaki, Y.~Nakamura,
\newblock \emph{Science Advances} \textbf{2017}, \emph{3} 425.

\bibitem{LQReview}
D.~Lachance-Quirion, Y.~Tabuchi, A.~Gloppe, K.~Usami, Y.~Nakamura,
\newblock \emph{Applied Physics Express} \textbf{2019}, \emph{12} 070101.

\bibitem{LQScience}
D.~Lachance-Quirion, S.~P. Wolski, Y.~Tabuchi, S.~Kono, K.~Usami, Y.~Nakamura,
\newblock \emph{Science} \textbf{2020}, \emph{367} 425.

\bibitem{Crescini}
N.~Crescini, C.~Braggio, G.~Carugno, A.~Ortolan, G.~Ruoso,
\newblock \emph{Appl. Phys. Lett.} \textbf{2020}, \emph{117} 144001.

\bibitem{LiberskySM}
M.~Libersky, R.~D. McKenzie, D.~M. Silevitch, P.~C.~E. Stamp, T.~F. Rosenbaum,
\newblock \emph{Phys. Rev. Lett.} \textbf{2021}, \emph{127} 207202.

\bibitem{StampGSM}
P.~C.~E. Stamp, D.~M. Silevitch, M.~Libersky, R.~D. McKenzie, A.~A. Geim, T.~F. Rosenbaum,
\newblock \emph{Phys. Rev. B} \textbf{2024}, \emph{110} 134420.

\bibitem{Hisatomi}
R.~Hisatomi, A.~Osada, Y.~Tabuchi, T.~Ishikawa, A.~Noguchi, R.~Yamazaki, K.~Usami, Y.~Nakamura,
\newblock \emph{Phys. Rev. B} \textbf{2016}, \emph{93} 174427.

\bibitem{Engelhardt}
F.~Engelhardt, V.~A. S.~V. Bittencourt, H.~Hueble, O.~Klein, S.~V. Kusminskiy,
\newblock \emph{Phys. Rev. Appl.} \textbf{2022}, \emph{18} 044059.

\bibitem{Puel}
T.~O. Puel, A.~T. Turflinger, S.~P. Horvath, J.~D. Thompson, M.~E. Flatt{\'e},
\newblock \emph{Phys. Rev. Research} \textbf{2025}, \emph{7} 033221.

\bibitem{Barbieri}
R.~Barbieri, C.~Braggio, G.~Carugno, C.~S. Gallo, A.~Lombardi, A.~Ortolan, R.~Pengo, G.~Ruoso, C.~C. Speake,
\newblock \emph{Physics of the Dark Universe} \textbf{2017}, \emph{15} 135.

\bibitem{CresciniDM}
N.~Crescini, D.~Alesini, C.~Braggio, G.~Carugno, D.~D. Gioacchino, C.~S. Gallo, U.~Gambardella, C.~Gatti, G.~Iannone, G.~Lamanna, C.~Ligi, A.~Lombardi, A.~Ortolan, S.~Pagano, R.~Pengo, G.~Ruoso, C.~C. Speake, L.~Taffarello,
\newblock \emph{Eur. Phys. J. C} \textbf{2018}, \emph{78} 1.

\bibitem{FlowerDM}
G.~Flower, J.~Bourhill, M.~Goryachev, M.~E. Tobar,
\newblock \emph{Physics of the Dark Universe} \textbf{2019}, \emph{25} 1.

\bibitem{Imamoglu}
A.~Imamo$\breve{\text{g}}$lu,
\newblock \emph{Phys. Rev. Lett.} \textbf{2009}, \emph{102} 083602.

\bibitem{Soykal}
O.~O. Soykal, M.~E. Flatt\'e,
\newblock \emph{Phys. Rev. Lett.} \textbf{2010}, \emph{104} 077202.

\bibitem{Huebl}
H.~Huebl, C.~W. Zollitsch, J.~Lotze, F.~Hocke, M.~Greifenstein, A.~Marx, R.~Gross, S.~T.~B. Goennenwein,
\newblock \emph{Phys. Rev. Lett.} \textbf{2013}, \emph{111} 127003.

\bibitem{Zhang}
X.~Zhang, C.~L. Zou, L.~Jiang, H.~X. Tang,
\newblock \emph{Phys. Rev. Lett.} \textbf{2014}, \emph{113} 156401.

\bibitem{Kubala}
B.~Kubala, J.~K{\:o}nig,
\newblock \emph{Phys. Rev. B} \textbf{2002}, \emph{65} 245301.

\bibitem{Heiss2004}
W.~D. Heiss,
\newblock \emph{J. Phys. A: Math. Gen.} \textbf{2004}, \emph{37} 2455.

\bibitem{Eleuch}
H.~Eleuch, I.~Rotter,
\newblock \emph{Acta Polytechnica} \textbf{2014}, \emph{54} 106.

\bibitem{Dhara2017}
S.~Dhara, C.~Chakraborty, K.~M. Goodfellow, L.~Qiu, T.~A. O'Loughlin, G.~W. Wicks, S.~Bhattacharjee, A.~N. Vamivakas,
\newblock \emph{Nature Physics} \textbf{2017}, \emph{14} 130.

\bibitem{HarderPRL}
M.~Harder, Y.~Yang, B.~M. Yao, C.~H. Yu, J.~W. Rao, Y.~S. Gui, R.~L. Stamps, C.~M. Hu,
\newblock \emph{Phys. Rev. Lett.} \textbf{2018}, \emph{121} 137203.

\bibitem{Harder2021}
M.~Harder, B.~M. Yao, Y.~S. Gui, C.~M. Hu,
\newblock \emph{J. Appl. Phys.} \textbf{2021}, \emph{129} 201101.

\bibitem{Bernier2014}
N.~R. Bernier, E.~G.~D. Torre, E.~Demler,
\newblock \emph{Phys. Rev. Lett.} \textbf{2014}, \emph{113} 065303.

\bibitem{Bernier}
N.~R. Bernier, L.~D. T{\'o}th, A.~K. Feofanov, T.~J. Kippenberg,
\newblock \emph{Phys. Rev. A} \textbf{2018}, \emph{98} 023841.

\bibitem{Yu}
W.~Yu, J.~Wang, H.~Y. Yuan, J.~Xiao,
\newblock \emph{Phys. Rev. Lett.} \textbf{2019}, \emph{123} 227201.

\bibitem{Proskurin2018}
I.~Proskurin, A.~S. Ovchinnikov, J.~I. Kishine, R.~L. Stamps,
\newblock \emph{Phys. Rev. B} \textbf{2018}, \emph{98} 220411.

\bibitem{Proskurin2019}
I.~Proskurin, R.~Mac{\^e}do, R.~L. Stamps,
\newblock \emph{New J. Phys.} \textbf{2019}, \emph{21} 1.

\bibitem{Proskurin2021}
I.~Proskurin, R.~L. Stamps,
\newblock \emph{Phys. Rev. B} \textbf{2021}, \emph{103} 195409.

\bibitem{Bhoi}
B.~Bhoi, B.~Kim, S.~H. Jang, J.~Kim, J.~yang, Y.~J. Cho, S.~K. Kim,
\newblock \emph{Phys. Rev. B} \textbf{2019}, \emph{99} 134426.

\bibitem{Yang2019}
Y.~Yang, J.~W. Rao, Y.~S. Gui, B.~M. Yao, W.~Lu, C.~M. Hu,
\newblock \emph{Phys. Rev. Applied.} \textbf{2019}, \emph{11} 054023.

\bibitem{Yang2020}
Y.~Yang, Y.-P. Wang, J.~W. Rao, Y.~S. Gui, B.~M. Yao, W.~Lu, C.-M. Hu,
\newblock \emph{Phys. Rev. Lett.} \textbf{2020}, \emph{125} 147202.

\bibitem{Xu}
P.~C. Xu, J.~W. Rao, Y.~S. Gui, X.~Jin, C.~M. Hu,
\newblock \emph{Phys. Rev. B} \textbf{2019}, \emph{100} 094415.

\bibitem{WangNonrecip}
Y.~P. Wang, J.~W. Rao, Y.~Yang, P.~C. Xu, Y.~S. Gui, B.~M. Yao, J.~Q. You, C.~M. Hu,
\newblock \emph{Phys. Rev. Lett.} \textbf{2019}, \emph{123} 127202.

\bibitem{Yao}
Y.~Yang, J.~W. Rao, Y.~S. Gui, B.~M. Yao, W.~Lu, C.~M. Hu,
\newblock \emph{Phys. Rev. Applied.} \textbf{2019}, \emph{11} 054023.

\bibitem{YuDissipativeCoupling}
C.~H. Yu, Y.~Yang, J.~W. Rao, P.~Hyde, Y.~P. Wang, B.~Zhang, Y.~S. Gui, C.~M. Hu,
\newblock \emph{AIP Advances} \textbf{2019}, \emph{9} 115012.

\bibitem{RaoTravelingPhotons}
J.~W. Rao, Y.~P. Wang, Y.~Yang, T.~Yu, Y.~S. Gui, X.~L. Fan, D.~S. Xue, C.~M. Hu,
\newblock \emph{Phys. Rev. B} \textbf{2020}, \emph{101} 064404.

\bibitem{WangHu}
Y.~P. Wang, C.~M. Hu,
\newblock \emph{J. Appl. Phys.} \textbf{2020}, \emph{127} 130901.

\bibitem{Xiao}
Y.~Xiao, H.~Wang, D.~Wang, R.~Lu, X.~Yan, H.~Guo, C.~M. Hu, K.~Xia, H.~Zhang,
\newblock \emph{Phys. Rev. B} \textbf{2021}, \emph{104} 115147.

\bibitem{Wang}
J.~Wang, J.~Xiao,
\newblock \emph{Phys. Rev. B} \textbf{2025}, \emph{111} 115425.

\bibitem{Lu}
C.~Lu, M.~Kim, Y.~Yang, Y.~S. Gui, C.~M. Hu,
\newblock \emph{J. Appl. Phys.} \textbf{2023}, \emph{134} 221101.

\bibitem{Hong}
Q.~Hong, W.~X. Wu, Y.~P. Peng, J.~Qian, C.~M. Hu, Y.~P. Wang,
\newblock \emph{Phys. Rev. A} \textbf{2024}, \emph{110} 052201.

\bibitem{RaoAntires}
J.~W. Rao, C.~H. Yu, Y.~T. Zhao, Y.~S. Gui, X.~L. Fan, D.~S. Xue, C.~M. Hu,
\newblock \emph{New Journal of Physics} \textbf{2019}, \emph{21} 1.

\bibitem{Hao}
Z.~Hao, Y.~Yao, K.~An, X.~Li, C.~Zhang, G.~Chai,
\newblock \emph{Phys. Rev. B} \textbf{2025}, \emph{112} 064416.

\bibitem{Ashida}
Y.~Ashida, Z.~Gong, M.~Ueda,
\newblock \emph{Advances in Physics} \textbf{2020}, \emph{69} 249.

\bibitem{Boventer1}
I.~Boventer, M.~Kl{\"a}ui, R.~Mac{\^e}do, M.~Weides,
\newblock \emph{New J. Phys.} \textbf{2019}, \emph{21} 125001.

\bibitem{Boventer2}
I.~Boventer, C.~D{\"o}rflinger, T.~Wolz, R.~Mac{\^e}do, R.~Lebrun, M.~Kl{\"a}ui, M.~Weides,
\newblock \emph{Phys. Rev. Research} \textbf{2020}, \emph{2} 013154.

\bibitem{grigoryan}
V.~L. Grigoryan, K.~Shen, K.~Xia,
\newblock \emph{Phys. Rev. B} \textbf{2018}, \emph{98} 024406.

\bibitem{rao2021interferometric}
J.~Rao, P.~Xu, Y.~Gui, Y.~Wang, Y.~Yang, B.~Yao, J.~Dietrich, G.~Bridges, X.~Fan, D.~Xue, et~al.,
\newblock \emph{Nature communications} \textbf{2021}, \emph{12}, 1 1933.

\bibitem{Bourcin}
G.~Bourcin, A.~Gardin, J.~Bourhill, V.~Vlaminck, V.~Castel,
\newblock \emph{Phys. Rev. Appl.} \textbf{2024}, \emph{22} 064036.

\bibitem{gardin2025level}
A.~Gardin, G.~Bourcin, C.~Person, C.~Fumeaux, R.~Lebrun, I.~Boventer, G.~C. Tettamanzi, V.~Castel,
\newblock \emph{Physical Review Applied} \textbf{2025}, \emph{23}, 1 014048.

\bibitem{Berry}
M.~V. Berry,
\newblock \emph{Czechoslovak Journal of Physics} \textbf{2004}, \emph{54} 1039.

\bibitem{Heiss}
W.~D. Heiss,
\newblock \emph{J. Phys. A: Math. Theor.} \textbf{2012}, \emph{45} 1.

\bibitem{zhang2017observation}
D.~Zhang, X.-Q. Luo, Y.-P. Wang, T.-F. Li, J.~You,
\newblock \emph{Nature communications} \textbf{2017}, \emph{8}, 1 1368.

\bibitem{Heiss1999}
W.~D. Heiss,
\newblock \emph{Eur. Phys. J. D} \textbf{1999}, \emph{7} 1.

\bibitem{Dembowski2001}
C.~Dembowski, H.~D. Gr{\:a}f, H.~L. Harney, A.~Heine, W.~D. Heiss, H.~Rehfeld, A.~Richter,
\newblock \emph{Phys. Rev. Lett.} \textbf{2001}, \emph{86} 787.

\bibitem{Dembowski2004}
C.~Dembowski, B.~Dietz, H.~D. Gr{\:a}f, H.~L. Harney, A.~Heine, W.~D. Heiss, A.~Richter,
\newblock \emph{Phys. Rev. E} \textbf{2004}, \emph{69} 056216.

\bibitem{Hernandez}
E.~Hern{\'a}ndez, A.~J{\'a}uregui, A.~Mondrag{\'o}n,
\newblock \emph{J. Phys. A: Math. Gen.} \textbf{2006}, \emph{39} 10087.

\bibitem{Lefebvre}
R.~Lefebvre, O.~Atabek, M.~{\v S}indelka, N.~Moiseyev,
\newblock \emph{Phys. Rev. Lett.} \textbf{2009}, \emph{103} 123003.

\bibitem{Uzdin}
R.~Uzdin, A.~Mailybaev, N.~Moiseyev,
\newblock \emph{J. Phys. A: Math. Theor.} \textbf{2011}, \emph{44} 1.

\bibitem{Milburn}
T.~J. Milburn, J.~Doppler, C.~A. Holmes, S.~Portolan, S.~Rotter, P.~Rable,
\newblock \emph{Phys. Rev. A} \textbf{2015}, \emph{92} 052214.

\bibitem{xu2016topological}
H.~Xu, D.~Mason, L.~Jiang, J.~Harris,
\newblock \emph{Nature} \textbf{2016}, \emph{537}, 7618 80.

\bibitem{Ghosh}
S.~N. Ghosh, Y.~D. Chong,
\newblock \emph{Scientific Reports} \textbf{2016}, \emph{6} 1.

\bibitem{Doppler}
J.~Doppler, A.~A. Mailybaev, J.~B{\:o}hm, U.~Kuhl, A.~Girschik, F.~Libisch, T.~J. Milburn, P.~Rabl, N.~Moiseyev, S.~Rotter,
\newblock \emph{Nature} \textbf{2016}, \emph{537} 1.

\bibitem{Schumer}
A.~Schumer, Y.~G.~N. Liu, J.~Leshin, L.~Ding, Y.~Alahmadi, A.~U. Hassan, H.~Nasari, S.~Rotter, D.~N. Christodoulides, P.~LiKamWa, M.~Khajavikhan,
\newblock \emph{Science} \textbf{2022}, \emph{375} 884.

\bibitem{Wojcik}
C.~C. Wojcik, K.~Wang, A.~Dutt, J.~Zhong, S.~Fan,
\newblock \emph{Phys. Rev. B} \textbf{2022}, \emph{106} 1.

\bibitem{Patil}
Y.~S.~S. Patil, J.~H{\:o}ller, P.~A. Henry, C.~Guria, Y.~Zhang, L.~Jiang, N.~Kralj, N.~Read, J.~G.~E. Harris,
\newblock \emph{Nature} \textbf{2022}, \emph{607} 271.

\bibitem{Chavva}
V.~Chavva, H.~Ribeiro,
\newblock \emph{arXiv:2501.07454v1} \textbf{2025}, 1--17.

\bibitem{rao2024braiding}
Z.~Rao, C.~Meng, Y.~Han, L.~Zhu, K.~Ding, Z.~An,
\newblock \emph{Nature Physics} \textbf{2024}, 1--8.

\bibitem{LambertModeSwitch}
N.~J. Lambert, A.~Schumer, J.~J. Longdell, S.~Rotter, H.~G.~L. Schwefel,
\newblock \emph{Nature Physics} \textbf{2025}, 1--9.

\bibitem{ZhangModeSwitch}
H.~Zhang, C.~Wang, Q.~Bao, Z.~Liu, H.~Liu, J.~J. Xiao,
\newblock \emph{Optics Letters} \textbf{2025}, \emph{50} 4154.

\bibitem{Wiersig2014}
J.~Wiersig,
\newblock \emph{Phys. Rev. Lett.} \textbf{2014}, \emph{112} 203901.

\bibitem{Wiersig2016}
J.~Wiersig,
\newblock \emph{Phys. Rev. A} \textbf{2016}, \emph{93} 033809.

\bibitem{chen2017exceptional}
W.~Chen, {\c{S}}.~Kaya~{\"O}zdemir, G.~Zhao, J.~Wiersig, L.~Yang,
\newblock \emph{Nature} \textbf{2017}, \emph{548}, 7666 192.

\bibitem{hodaei2017enhanced}
H.~Hodaei, A.~U. Hassan, S.~Wittek, H.~Garcia-Gracia, R.~El-Ganainy, D.~N. Christodoulides, M.~Khajavikhan,
\newblock \emph{Nature} \textbf{2017}, \emph{548}, 7666 187.

\bibitem{CaoSensing}
Y.~Cao, P.~Yan,
\newblock \emph{Phys. Rev. B} \textbf{2019}, \emph{99} 214415.

\bibitem{Zhong}
Q.~Zhong, J.~Ren, M.~Khajavikhan, D.~N. Christodoulides, S.~K. {\:O}zdemir, R.~El-Ganainy,
\newblock \emph{Phys. Rev. Lett.} \textbf{2019}, \emph{122} 153902.

\bibitem{chen2019}
C.~Chen, L.~Jin, R.~B. Liu,
\newblock \emph{New Journal of Physics} \textbf{2019}, \emph{21} 1.

\bibitem{ZhangSensing}
M.~Zhang, W.~Sweeney, C.~W. Hsu, L.~Yang, A.~D. Stone, L.~Jiang,
\newblock \emph{Phys. Rev. Lett.} \textbf{2019}, \emph{123} 180501.

\bibitem{WiersigReview}
J.~Wiersig,
\newblock \emph{Photonics Research} \textbf{2020}, \emph{8} 1457.

\bibitem{LiStochasticEP}
Z.~Li, C.~Li, Z.~Xiong, G.~Xu, Y.~R. Wang, X.~Tian, X.~Yang, Z.~Liu, Q.~Zeng, R.~Lin, Y.~Li, J.~K.~W. Lee, J.~S. Ho, C.-W. Qiu,
\newblock \emph{Phys. Rev. Lett.} \textbf{2023}, \emph{130} 227201.

\bibitem{WangCPAEP}
C.~Wang, W.~R. Sweeney, A.~D. Stone, L.~Yang,
\newblock \emph{Science} \textbf{2021}, \emph{373} 1.

\bibitem{chongCPA}
Y.~D. Chong, L.~Ge, H.~Cao, A.~D. Stone,
\newblock \emph{Phys. Rev. Lett.} \textbf{2010}, \emph{105}, 053901 1.

\bibitem{Baranov}
D.~G. Baranov, A.~Krasnok, T.~Shegai, A.~Al{\`u}, Y.~Chong,
\newblock \emph{Nature Reviews: Materials} \textbf{2017}, \emph{2} 17064.

\bibitem{guo2023singular}
C.~Guo, J.~Li, M.~Xiao, S.~Fan,
\newblock \emph{Physical Review B} \textbf{2023}, \emph{108}, 15 155418.

\bibitem{SweeneyRSM}
W.~R. Sweeney, C.~W. Hsu, A.~D. Stone,
\newblock \emph{Phys. Rev. A} \textbf{2020}, \emph{102} 1.

\bibitem{hsiehOpenrings}
L.-H. Hsieh, K.~Chang,
\newblock \emph{IEEE Transactions on Microwave Theory and Techniques} \textbf{2002}, \emph{50}, 2 453.

\bibitem{GardinerZollerBook}
C.~W. Gardiner, P.~Zoller,
\newblock \emph{{Quantum Noise}},
\newblock Springer-Verlag, 3rd edition, \textbf{2004}.

\bibitem{WallsMilburn}
D.~F. Walls, G.~J. Milburn,
\newblock \emph{Quantum Optics},
\newblock Springer, \textbf{2020}.

\bibitem{hyde2016indirect}
P.~Hyde, L.~Bai, M.~Harder, C.~Match, C.-M. Hu,
\newblock \emph{Applied Physics Letters} \textbf{2016}, \emph{109}, 15.

\bibitem{MetelmannClerk}
A.~Metelmann, A.~A. Clerk,
\newblock \emph{Phys. Rev. X} \textbf{2015}, \emph{5} 021025.

\bibitem{GardinerCollett}
C.~W. Gardiner, M.~J. Collett,
\newblock \emph{Phys. Rev. A} \textbf{1985}, \emph{31} 3761.

\bibitem{comsol}
{{COMSOL AB}},
\newblock {COMSOL Multiphysics}, \textbf{2024},
\newblock \urlprefix\url{https://www.comsol.com}.

\bibitem{Miri}
M.~A. Miri, A.~Al{\`u},
\newblock \emph{Science} \textbf{2019}, \emph{363} 1.

\bibitem{Bender}
C.~M. Bender, S.~Boettcher,
\newblock \emph{Phys. Rev. Lett.} \textbf{1998}, \emph{80} 5243.

\bibitem{Ozdemir}
S.~K. {\"Ozdemir}, S.~Rotter, F.~Nori, L.~Yang,
\newblock \emph{Nature Materials} \textbf{2019}, \emph{18} 783.

\bibitem{ZhangPT}
Z.~Zhang, C.~Xin, H.~Liu,
\newblock \emph{Advanced Electronic Materials} \textbf{2023}, \emph{10} 2300674.

\bibitem{Pozar}
D.~M. Pozar,
\newblock \emph{Microwave Engineering},
\newblock John Wiley and Sons, \textbf{2005}.

\bibitem{ZhangEP3}
X.~Zhang, K.~Ding, X.~Zhou, J.~Xu, D.~Jin,
\newblock \emph{Phys. Rev. Lett.} \textbf{2019}, \emph{123} 237202.

\bibitem{mao2024exceptional}
W.~Mao, Z.~Fu, Y.~Li, F.~Li, L.~Yang,
\newblock \emph{Science Advances} \textbf{2024}, \emph{10}, 14 eadl5037.

\bibitem{LeeNonlinear}
O.~Lee, K.~Yamamoto, M.~Umeda, C.~W. Zollitsch, M.~Elyasi, T.~Kikkawa, E.~Saitoh, G.~E.~W. Bauer, H.~Kurebayashi,
\newblock \emph{Phys. Rev. Lett.} \textbf{2023}, \emph{130} 046703.

\bibitem{Clerk}
A.~A. Clerk, M.~H. Devoret, S.~M. Girvin, F.~Marquardt, R.~J. Schoelkopf,
\newblock \emph{Reviews of Modern Physics} \textbf{2010}, \emph{82} 1155.

\end{thebibliography}





\maketitlesup

\author{Mawgan A. Smith$~^{1,*}$}
\author{Ryan D. McKenzie$~^{2,*}$}
\author{Alban Joseph$~^1$}
\author{Robert L. Stamps$~^2$}
\author{Rair Mac\^edo$~^1$}

\begin{affiliations}

$^1$James Watt School of Engineering, Electronics \& Nanoscale Engineering Division, University of Glasgow, Glasgow G12 8QQ, United Kingdom\\

$^2$Department of Physics and Astronomy, University of Manitoba, 30A Sifton Road, Winnipeg, R3T 2N2, Manitoba, Canada

$^*$\emph{These authors contributed equally to this work.}

\end{affiliations}

\section{Experimental Methods and Charaterisation}
\label{ap:methods}

\subsection{Experimental Setup and Resonator Design}
\label{ap:design}

Measurements were performed using a Anritsu Vector Star MS4642B, where the output power is set to 10 dBm across all measurements. The output is fed through a 50 $\Omega$ coaxial cable to a balanced power splitter. The relative phase is controlled using an Vaunix LPS-802 and the relative amplitude controlled with a Vaunix LDA-203B digital attenuator. Port four of the resonator has a 50 $\Omega$ termination at its output to ensure that there are no signal reflections back into the resonator. 

\begin{figure}
	\centering    
    \mbox{\includegraphics[width=0.62\linewidth]{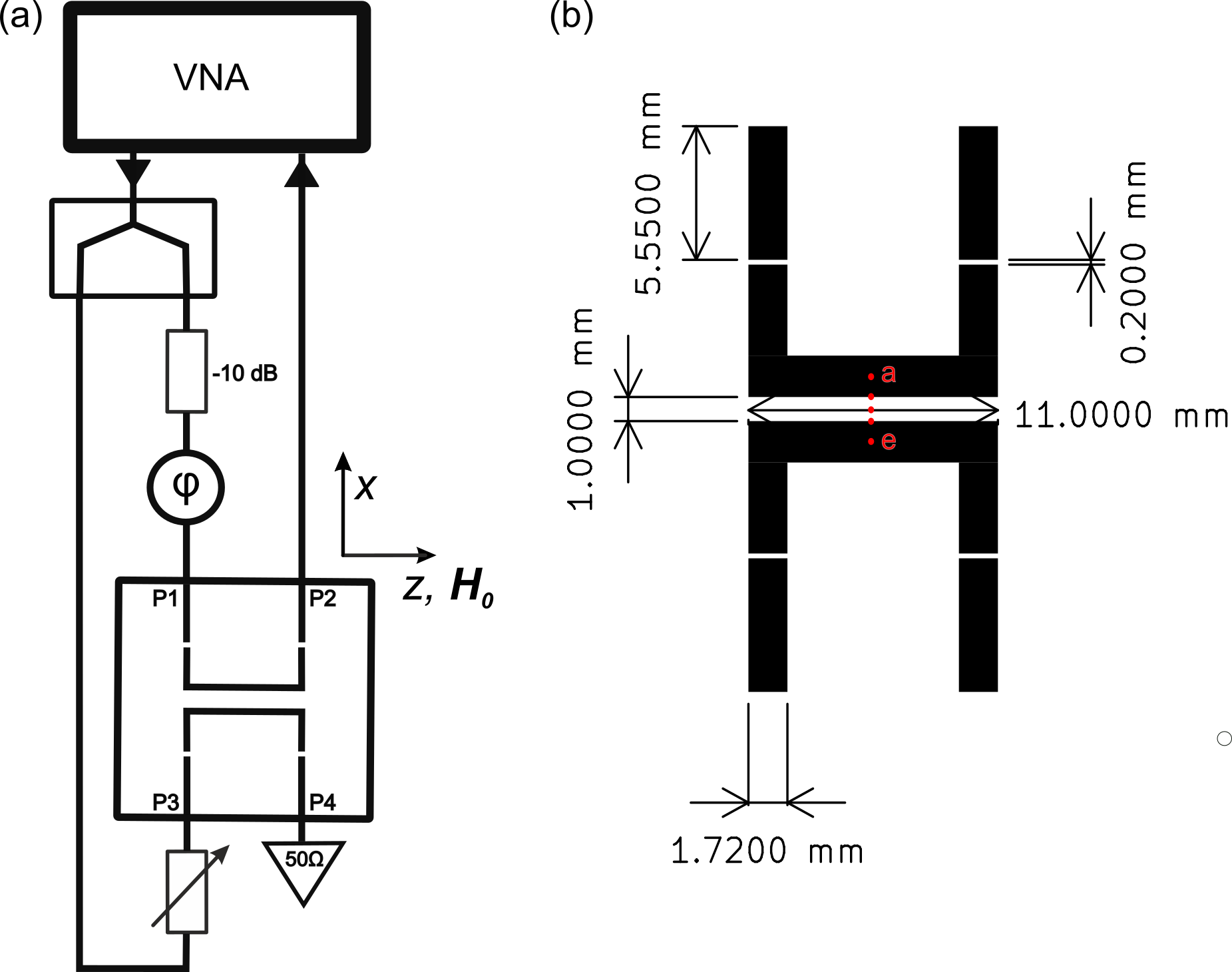}}   
    \caption{(a) Schematic of experimental setup. The vector network analyzer outputs to a balanced power splitter which is fed into P1 one and P3 of the resonator. The path into port one has static attenuation and phase control of the signal going into the resonator port, this controls the relative phase. The path into port three is controlled by a variable attenuator to set the relative amplitude. (b) Dimensional drawing of the planar resonator tracks. Additional labels in red highlight the sample positions ($a-e$) with point c at the midpoint between the two cavities. If we consider the midpoint $c$ as the origin then the sample positions are approximately [1.36, 0.5, 0., -0.5, -1.36 ]~mm though we note there may be some deviation from this due to parallax error in placing the sample.} 
    \label{fig:dimensions}
\end{figure}

The experimental setup is detailed in Fig. \ref{fig:dimensions} (a). Before taking the measurements, the phase and amplitude at the end of the signal paths going into the resonator ports must be determined so that the relative phase and amplitude can be set accurately. The loss through path one (into P1) containing the phase shifter is determined to be $\sim$ -17 dB and the phase of the signal (with the LPS set to 0$^{\circ}$) is $\sim$ 13$^{\circ}$. For path three (into P3) the unattenuated signal is $\sim$ -10 dB, which is then set to match specific amplitudes relative to path one, and the phase at the end of this signal path is $\sim$ -14$^{\circ}$. From this we can determine the necessary phase setting to achieve desired relative phases, allowing us to construct the phase maps of the antiresonance points.

The resonator used in the proof-of-principle experiment is fabricated on copper-plated RO4350B, with a dielectric thickness of 0.76 mm, copper thickness of 35 $\mu$m, relative permittivity of $\epsilon_{r}=3.48$, and dielectric loss tangent of tan($\delta$) = 0.0037. The width of the copper tracks of both the feedlines and the resonator are set to 1.72 mm to ensure correct impedance matching (50 $\Omega$) to the SMA connectors which connect the resonator to the external circuitry. The width of the individual resonators is 11 mm and the height 5.5 mm, forming an open square. The distance between the two resonators, which determines the inductive coupling $J$ is set to 1 mm. The coupling gaps between the resonators and the feedlines is 0.2 mm, this determines the external gain/loss of the resonator system, contributing to the total Q-factor of the modes. A dimensional drawing is presented in Fig. \ref{fig:dimensions} (b). We use a 0.2 mm diameter commerical YIG sphere from Ferrisphere Inc. in the experiments.

\newpage

\subsection{Parameter Estimation}
\label{ap:estimation}

\begin{figure}[!h]
\centering
\includegraphics[width=0.9\linewidth]{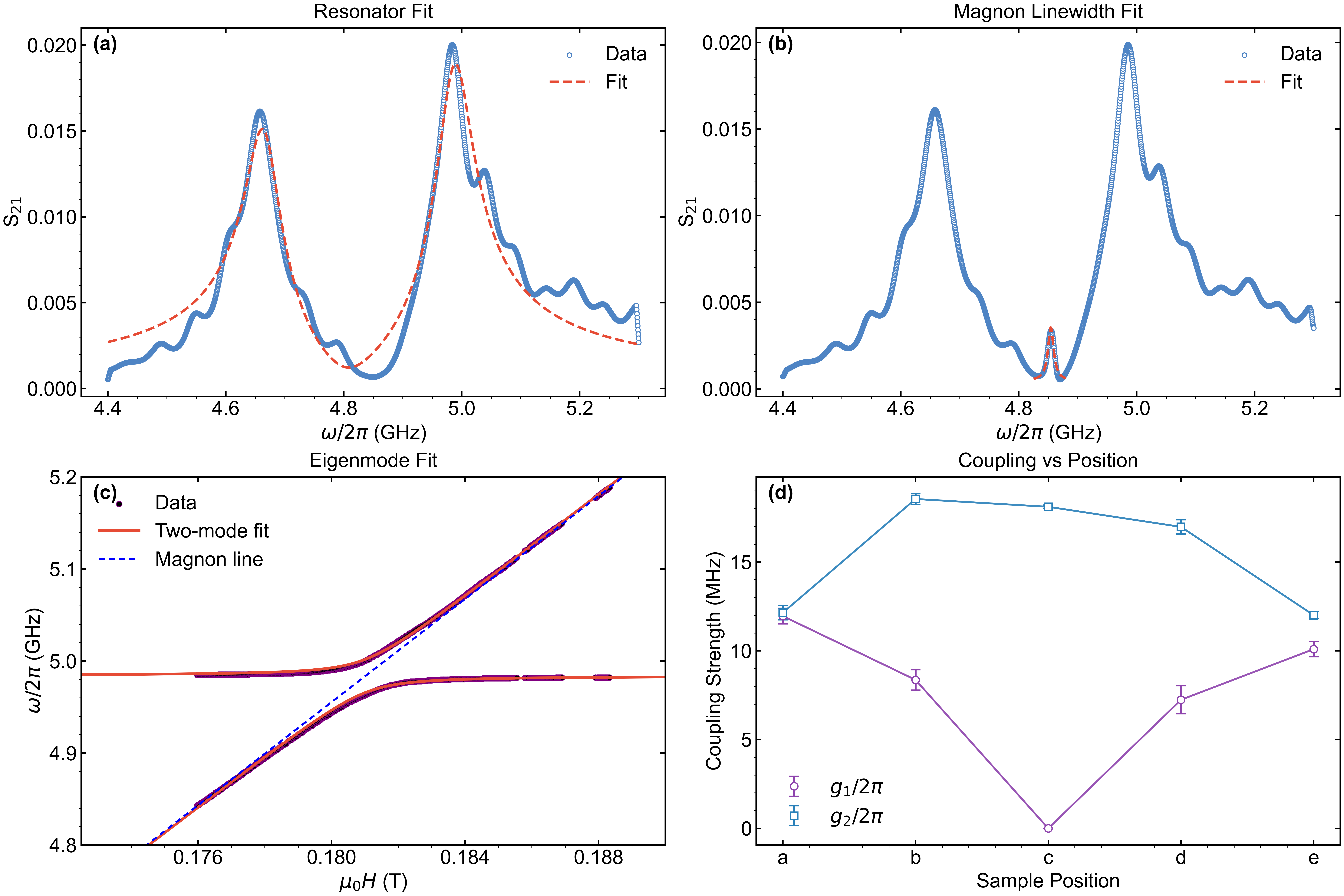}
\captionof{figure}{(a) Direct transmission from port one to two of the resonator with the magnon mode far off resonance. The spectrum is fit to the direct transmission [Eq. (\ref{eq:tdti})] to determine the resonator parameters. (b) Direct transmission spectrum with the magnon mode present but de-tuned from the dual cavity resonances. This is used to estimate the magnon linewidth by fitting a Lorentzian function to the associated data points \cite{LeeNonlinear}. (c) Example of eigenmode fitting used to determine coupling strength. When the sphere is in the centre of the two resonators we can use a simple two mode fit. The obtained coupling strength does not change when the extra input is switched on at port three. (d) Change in coupling strength to the lower frequency (g${_1}$) and upper frequency (g$_{2}$) modes, moving from the centre of the upper resonator at point $a$ through the centre of the two resonators $c$ to the centre of the lower resonator $e$, as seen in Fig.\ref{fig:dimensions} (b).}
\label{fig:combined}
\end{figure}

Here, we discuss the fittings used to estimate the system parameters. Of particular note in the fitting of the cavity modes is the shift of the minimum in the fit, corresponding to $\omega_{a}$, compared to the experimental data. In the model $\omega_{a}$ is the natural frequency of a single cavity and is identical for both, this should also correspond to the antiresonance frequency in the one-drive case. The cause of the discrepancy is likely due to two factors: absence of the frequency dependent linewidths in the model, and slight differences between the natural frequencies of each of the cavities due to limits in fabrication tolerance causing imperfections. Despite this, the fitted parameters obtained lead to good agreement between experiment and theory as seen in Fig. \ref{fig:combined}.

To estimate the magnon dissipation, we fit a Lorentzian to the magnon mode when it is detuned from the resonator modes. The obtained FWHM of 11.65 MHz is larger than typical for polished YIG spheres in this frequency range, we attribute this to the proximity of the sample to the copper tracks leading to additional extrinsic radiative damping which is larger than the intrinsic losses of the YIG in this case.

In Fig. \ref{fig:combined} (c) we show examples of the eigenmode fits to the peaks from the magnon-polariton modes. These are used to obtain the coherent magnon-photon coupling strengths for different positions of the sample. The values of the couplings to both cavity modes $\omega_{1,2}$, at the sample positions shown in Fig. \ref{fig:dimensions} (b), are shown in Fig. \ref{fig:combined} (d). 

\newpage

\subsection{Calibration}
\label{ap:calibration}
\begin{figure}[!h]
\centering
\includegraphics[width=0.9\linewidth]{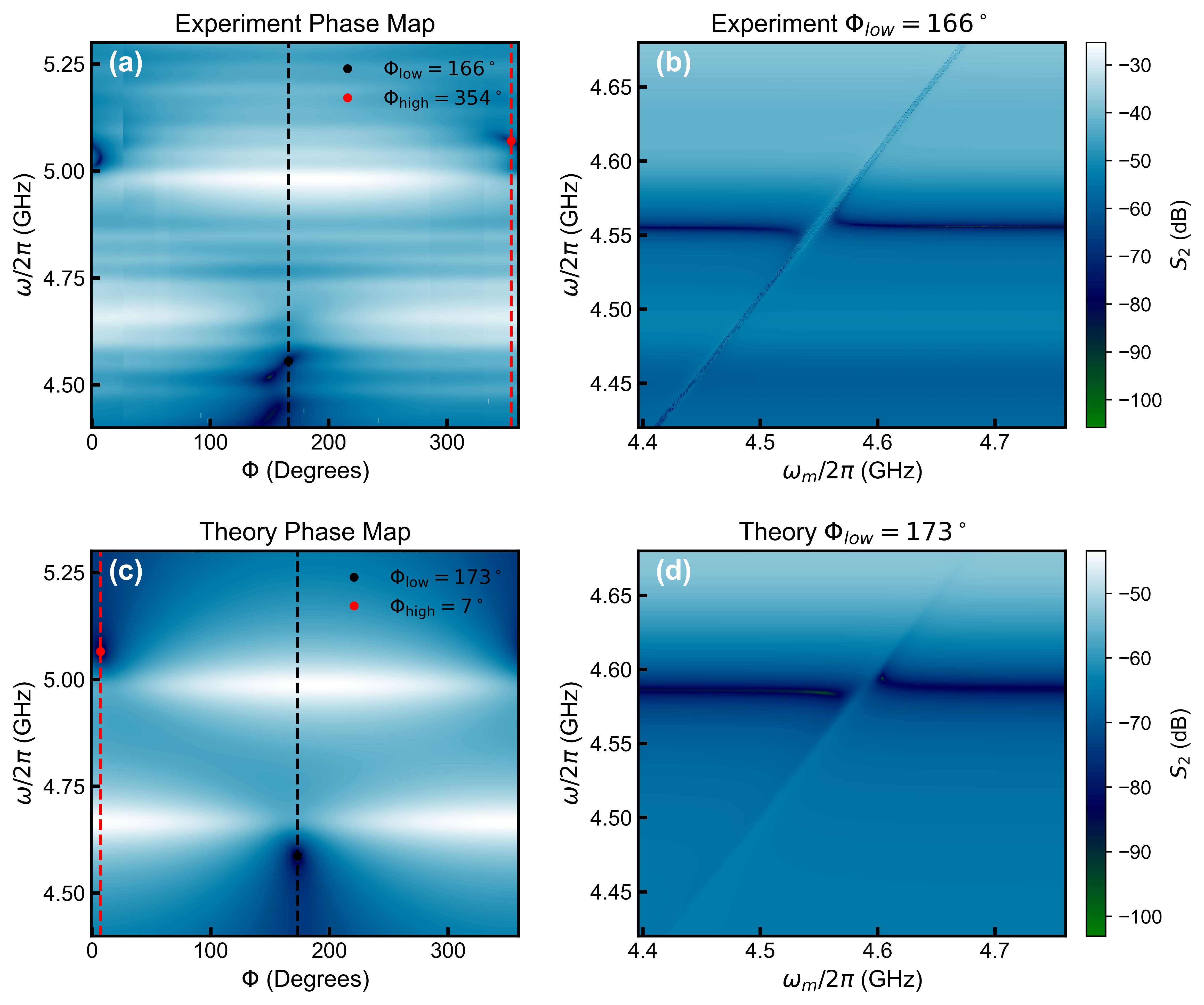}
\captionof{figure}{Phase calibration map for a fixed relative amplitude $\delta = 1.5$. (a) Experimental and (c) theoretical phase maps with annotated best $\Phi$. (b) Experimental and (d) theoretical antimode level attraction spectra using the calibrated phase value. }
\label{fig:phase_LA}
\end{figure}

In the experiment, we are interested in the behaviour of the antimodes. We focus our attention on three relative amplitude values $\delta = 0.5, 1.0, 1.5$ to highlight the transition to level attraction of the antimodes when $\delta > 1$. The antiresonance condition in the resonator system is achieved for particular combinations of phase and amplitude. We construct a phase calibration map to locate the relevant phase value as shown in Fig. \ref{fig:phase_LA} (a). 

Upon comparing the experimental data to the theoretically calculated phase map, we note that the minima occur at slightly different values, $\Phi_{exp} = 166^{\circ}$ and $\Phi_{the} = 173^{\circ}$ at frequencies $\omega/2\pi = 4.556$ GHz and $\omega/2\pi = 4.587$ GHz, respectively. Knowing the theoretical relative phase and amplitude, we can compare the experimental observation of level attraction to that calculated from input-output theory. There is a slight increase in frequency of the antimode in the theoretical spectrum due to the different phase calibration, but the main features of the experimental data agree well with the theory.

\newpage

\section{Input-Output Calculation}
\label{ap:IO}

We consider the system Hamiltonian in the cavity basis given by Eq. (\ref{eq:Hsys}). Each of the four resonator ports is coupled to a bath with the system bath interaction given by Eq. (\ref{eq:Hint}). Using input-output theory \cite{GardinerCollett, GardinerZollerBook, WallsMilburn}, we derive equations of motion for the system and bath operators. The result is used to determine the S-matrix.

The bath Hamiltonian and system-bath interaction are defined in the continuum limit. Note that
\begin{align}
&\mathcal{H} =\sum_q \hbar \omega_q b_q^{\dag} b_q 
\\ \nonumber &\qquad\Rightarrow\qquad
\mathcal{H} = \hbar\int d\omega \ \omega \rho(\omega) b_{\omega}^{\dag} b_{\omega}
\end{align}
where $\rho(\omega) = \sum_q \delta(\omega-\omega_q)$ is the density of states. The density of states is discussed in Ref. \cite{Clerk} (see Supplemental Information~E2). Here, as is common practice, we assume the bath Hamiltonian is defined by Eq. (\ref{eq:Hbath}); the operators $b_n(t,\omega)$ then have units of $1/\sqrt{\text{frequency}}$ and the system-bath couplings have units of $\sqrt{\text{frequency}}$.

The Heisenberg equations of motion of the bath operators are
\begin{align}
\dot{\boldsymbol{b}} = -i\omega \boldsymbol{b} + K^{\dag} \boldsymbol{a}.
\end{align}
Recall that $\boldsymbol{b}^T=(b_1,b_2,b_3,b_4)$ are the bath annihilation operators and $\boldsymbol{a}^T=(a_1,a_2)$ are annihilation operators for the cavity photons. The matrix elements of $K^{\dag}$ may be read from Eq. (\ref{eq:couplings}) of Sec. \ref{sec:system}. One may integrate these equations to obtain the input equation ($t_0<t$)
\begin{align}
\int d\omega \ \boldsymbol{b}(t,\omega) &= \sqrt{2\pi} \boldsymbol{b}^{IN}(t) 
+ \pi K^{\dag} \boldsymbol{a}(t),
\end{align}
and the output equation ($t<t_f$)
\begin{align}
\int d\omega \ \boldsymbol{b}(t,\omega) &= \sqrt{2\pi} \boldsymbol{b}^{OUT}(t) 
- \pi K^{\dag} \boldsymbol{a}(t),
\end{align}
where the inputs and outputs are defined by
\begin{align}
\label{eq:InOut}
b_{n}^{IN}(t) &=  \frac{1}{\sqrt{2\pi}} \int d\omega \  
e^{-i\omega(t-t_0)} b_n(t_0,\omega) 
\\ \nonumber
b_{n}^{OUT}(t) &=  \frac{1}{\sqrt{2\pi}} \int d\omega \  e^{-i\omega(t-t_f)} b_n(t_f,\omega).
\end{align}
At time $t$, the system is subject to the boundary conditions
\begin{align}
\label{eq:bcs}
\boldsymbol{b}^{OUT}-\boldsymbol{b}^{IN} = \sqrt{2\pi} K^{\dag} \boldsymbol{a}
\end{align}
Having obtained equations of motion of the bath operators, we now consider the system operators.

The dynamics of the system operators follow from
\begin{align}
\dot{a_j} &= -\frac{i}{\hbar}[a_j,\mathcal{H}_{sys}] 
- \sum_{n=1}^{4} K_{jn} \int d\omega \ b_n (\omega)
\\ \nonumber
\dot{m} &= -\frac{i}{\hbar}[m,\mathcal{H}_{sys}]. 
\end{align}
The magnon modes do not couple directly to the bath modes and the equation of motion is simply
\begin{align}
\dot{m} = -i\omega_m m -i \sum_{j=1}^2 g_j a_j.
\end{align}
In frequency space one finds $m=S_m(g_1 a_1+g_2a_2)$ where
\begin{align}
S_m=\frac{1}{\omega-\omega_m+i\gamma_m/2}
\end{align}
determines the response of the magnons. The phenomenological damping factor $\gamma_m$ is included to account for intrinsic damping of the magnon mode. One may make use of the equations of motion of the bath operators and the magnon mode to determine the equations of motion of the cavity modes.

In frequency space, one finds the equations of motion of the cavity modes to be

\vspace{-0.5cm}
\begin{align}
\label{eq:inout}
-i\omega a_1 &= -i\omega_1 a_1 - i g_1^2 S_m a_1 - i g_1 g_2 S_m  a_2
- \frac{\gamma_{\kappa}}{2} a_1 - g_a a_2 
- \sqrt{2\pi} \sum_{n=1}^4 K_{1n} b_n^{IN}
\qquad\quad\  t_0 < t
\\ \nonumber 
-i\omega a_1 &= -i\omega_1 a_1 - i g_1^2 S_m  a_1 - i g_1 g_2 S_m  a_2
+ \frac{\gamma_{\kappa}}{2} a_1 + g_a a_2 
- \sqrt{2\pi} \sum_{n=1}^4 K_{1n} b_n^{OUT} 
\qquad\ \  t < t_f
\\ \nonumber 
-i\omega a_2 &= -i\omega_2 a_2 - i g_2 g_1 S_m  a_1 - i g_2^2 S_m  a_2 
- g_a a_1 - \frac{\gamma_{\kappa}}{2} a_2  
- \sqrt{2\pi} \sum_{n=1}^4 K_{2n} b_n^{IN} 
\qquad\quad\  t_0 < t
\\ \nonumber
-i\omega a_2 &= -i\omega_2 a_2 - i g_2 g_1 S_m  a_1 - i g_2^2 S_m  a_2 
+ g_a a_1 + \frac{\gamma_{\kappa}}{2} a_2 
- \sqrt{2\pi} \sum_{n=1}^4 K_{2n} b_n^{OUT} 
\qquad\ \  t < t_f
\end{align}
\vspace{0.2cm}

\noindent
The damping parameter $\gamma_{\kappa}$ and the dissipative cavity mode coupling $g_a$ are defined by
\begin{align}
\left[ \begin{array}{cc}
\gamma_{\kappa} & 2g_a  \\ 2g_a & \gamma_{\kappa}   \\
\end{array} \right] = 2\pi K K^{\dag}.
\end{align}
The damping parameter represents the damping of the modes due to their couplings to the four resonator ports. The two cavity modes share a common coupling to each of the baths; this gives rise to the dissipative coupling $g_a$ \cite{MetelmannClerk, WangHu}. One finds
\begin{align}
g_a = \frac{\pi}{2} [\kappa_{u1}^2 + \kappa_{u2}^2 - \kappa_{l3}^2 - \kappa_{l4}^2],
\end{align}
with the minus signs stemming from the antisymmetrisation of the low-frequency cavity mode. As noted in the introduction, if the couplings at the resonator ports are assumed to be equal then $g_a=0$. In what follows, we will assume this is the case ($\kappa_{un}=\kappa_{ln}=\kappa$); the damping due to the cavity ports is then $\gamma_{\kappa}=4\pi \kappa^2$.

The cavity mode equations may be simplified by introducing the functions
\begin{align}
S_{jm}^{-1} = \omega-\omega_j +i\frac{\gamma_j}{2}-\frac{g_j^2}{\omega-\omega_m+i\gamma_m/2}.
\end{align}
Here, each cavity mode is coupled to the magnon mode. Near resonance, $\omega_j \approx \omega_m$, the poles of $S_{jm}$ are magnon-polariton modes. The cavity mode dampings are defined by $\gamma_{j}=\gamma_{\kappa}+\gamma_{int,j}$, where $\gamma_{int,j}$ accounts for intrinsic damping. 

Similar to Ref. \cite{Bourcin}, we express the equations of motion in matrix form. In terms of $S_{jm}$, with equal couplings at the cavity ports ($g_a=0$), one finds the input equations to be ($t_0<t$)
\begin{align}
\label{eq:EOMcavity}
\Omega \boldsymbol{a} = -i \sqrt{2\pi} K \boldsymbol{b}^{IN}
\end{align}
with
\begin{align}
\Omega = \left[ \begin{array}{cc}
S_{1m}^{-1} & - g_{12} \\ -g_{21} & S_{2m}^{-1}  \\
\end{array} \right].
\end{align}
The frequency-dependent coupling between the high and low-frequency magnon-polariton modes is given by
\begin{align}
g_{12}(\omega)=g_{21}(\omega) = g_1g_2S_{m}(\omega).
\end{align}
This coupling is strongest if $\omega \approx \omega_m$ where the magnon mode is able to mediate interactions between the two cavity modes. 

One may also use Eq. (\ref{eq:inout}) to obtain an output equation 
\begin{align}
\label{eq:OUT}
[\Omega - i 2\pi KK^{\dag}] \boldsymbol{a} = -i \sqrt{2\pi} K \boldsymbol{b}^{OUT}.
\end{align}
In the output equation, loss at the cavity ports becomes gain from the drive field (the sign of $\gamma_{\kappa}$ changes).

If $\text{det}[\Omega] \neq 0$, one may invert the time evolution matrix to obtain 
\begin{align}
\label{eq:IN}
\boldsymbol{a} = -i \sqrt{2\pi} \Omega^{-1} K \boldsymbol{b}^{IN}.
\end{align}
Defining
\begin{align}
S_c^{-2} \equiv \text{det}[\Omega] = S_{1m}^{-1}S_{2m}^{-1} - g_{12}^2,
\end{align}
the inverse of the time evolution matrix is
\begin{align}
\Omega^{-1} = \left[ \begin{array}{cc}
S_{a_1} &  g_{12} S_c^2 \\ g_{21} S_c^2  & S_{a_2}  \\
\end{array} \right],
\end{align}
where
\begin{align}
S_{a_1} = \frac{S_{1m}}{1 - g_{12}^2 S_{1m} S_{2m}},
\end{align}
and similarly for $S_{a_2}$. The function $S_{a_1}$, for example, describes a feedback loop in which the low frequency magnon-polaritons ($S_{1m}$) receive feedback from the high frequency magnon-polaritons ($S_{2m}$) mediated by the magnons ($g_{12}$).

From Eqs. (\ref{eq:OUT}) and (\ref{eq:IN}) the input-output relations are
\begin{align}
\label{eq:Somega}
K b^{OUT} = [\Omega - i 2\pi KK^{\dag}] \Omega^{-1} K \boldsymbol{b}^{IN}
\end{align}
or, making use of the boundary conditions [Eq. (\ref{eq:bcs})], one finds
\begin{align}
\boldsymbol{b}^{OUT} = (\mathbbm{1}-i2\pi K^{\dag} \Omega^{-1} K) \boldsymbol{b}^{IN}
\end{align}
The term in brackets, $S=\mathbbm{1}-i2\pi K^{\dag} \Omega^{-1} K$ is the S-matrix of the four-port network; its elements $S_{mn}$ yield the output at port $m$ for a given input at port $n$. We adopt a phase convention in which the inputs and outputs are defined by
\begin{align}
\label{eq:x}
x_{1,4}^{IN} &= b_{1,4}^{IN} \qquad\qquad x_{1,4}^{OUT} = -b_{1,4}^{OUT}
\\ \nonumber
x_{2,3}^{IN} &= -b_{2,3}^{IN} \quad\qquad\ x_{2,3}^{OUT} = b_{2,3}^{OUT}.
\end{align}
This will flip the sign of certain S-matrix elements. This convention has been chosen so that the inputs and outputs in the theory match the signals to and from the vector network analyzer used in the experiments.

\section{Empty Resonators}
\label{ap:IOempty}

In the absence of the YIG sphere, the system Hamiltonian can be written as 
\begin{align}
\label{eq:Hsys2}
\frac{\mathcal{H}_{\text{sys}}}{\hbar} = \omega_a &a_u^{\dag} a_u + \omega_a a_l^{\dag} a_l + J (a_u^{\dag} a_l + a_u a_l^{\dag}).
\end{align}
Here, we make use of input-output theory to obtain equations of motion of the photon operators when the upper resonator is driven, and when both resonators are driven. This allows for reflectionless scattering modes and coherent perfect absorption, as discussed in Sec. \ref{sec:CPAandE}.

Making use of input-output theory, in the four-port system under consideration, the input equations are
\begin{align}
i\left[ \begin{array}{c}
\dot{a}_u \\ \dot{a}_l
\end{array} \right] = h_{res}
\left[ \begin{array}{c}
a_u \\ a_l
\end{array} \right] - i\sqrt{2\pi} K_0 \boldsymbol{b}^{IN},
\end{align}
where 
\begin{align}
h_{res} = \left[ \begin{array}{cc}
\omega_a-i\gamma_u/2 & J  \\ 
J  & \omega_a-i\gamma_l/2  \\
\end{array} \right].
\end{align}
The coupling matrix $K_0$ is defined in Eq. (\ref{eq:K0}), and the inputs are defined in Eq. (\ref{eq:InOut}) of Supplemental Information~\ref{ap:IO}. The damping parameters, 
\begin{align}
\label{eq:dampings}
\gamma_u =  \gamma_{u,\kappa} + \gamma_{u,int} \quad\text{and}\quad
\gamma_l = \gamma_{u,\kappa} + \gamma_{l,int},
\end{align}
include damping from the resonator ports, $\gamma_{u,\kappa}=2\pi (\kappa_{u1}^2+\kappa_{u2}^2)$ and $\gamma_{l,\kappa}=2\pi (\kappa_{l3}^2+\kappa_{l4}^2)$, as well as intrinsic damping of the resonator photons. 

The non-Hermitian Hamiltonian governing the system dynamics is 
\begin{align}
\frac{\mathcal{H}_{res}}{\hbar} = \left[ \begin{array}{cc}
a_u^{\dag} & a_l^{\dag} \end{array} \right] h_{res}
\left[ \begin{array}{c}
a_u \\ a_l \end{array} \right].
\end{align}
The eigenvalues of $h_{res}$, 
\begin{align}
&z_{\pm} = \frac{z_u+z_l}{2} \pm \frac{1}{2} \sqrt{(z_u-z_l)^2+4J^2} \\ \nonumber
&\text{with}\qquad z_{u,l} = \omega_{a} - i\gamma_{u,l}/2,
\end{align}
determine the modes, or resonances, of the system (the poles of the S-matrix). In the absence of damping $z_{\pm}$ reduces to $\omega_{2,1}$ [see Eq. (\ref{eq:Hsys})].

If resonator ports one and two are driven the output equation is
\begin{align}
i\left[ \begin{array}{c}
\dot{a}_u \\ \dot{a}_l
\end{array} \right] = h_{RSM}
\left[ \begin{array}{c}
a_u \\ a_l
\end{array} \right] - i\sqrt{2\pi} K_0 \boldsymbol{b}^{OUT},
\end{align}
where 
\begin{align}
h_{RSM} = \left[ \begin{array}{cc}
\omega_a+i\widetilde{\gamma}_u/2 & J  \\ 
J  & \omega_a-i\gamma_l/2  \\
\end{array} \right]
\end{align}
and $\widetilde{\gamma}_u = \gamma_{u,\kappa} - \gamma_{u,int}$. The upper resonator now exhibits gain from the drive field, rather than the loss from the resonator ports seen in the input equation. If gain from the drive field is balanced by intrinsic system losses and loss from the lowering scattering channels one obtains reflectionless scattering modes.

When all four resonator ports are driven, rather than $h_{RSM}$ in the output equation, one has
\begin{align}
h_{ar} = \left[ \begin{array}{cc}
\omega_a+i\widetilde{\gamma}_u/2 & J  \\ 
J  & \omega_a+i\widetilde{\gamma}_l/2  \\
\end{array} \right],
\end{align}
where $\widetilde{\gamma}_u$ is as before, and $\widetilde{\gamma}_{l} = \gamma_{l,\kappa} - \gamma_{l,int}$. The system now exhibits gain from all four resonator ports. Defining $\widetilde{z}_{u,l} = \omega_a + i \widetilde{\gamma}_{u,l}/2$, the eigenvalues of $h_{ar}$ are
\begin{align}
z_{ar,\pm} = \frac{\widetilde{z}_u+\widetilde{z}_l}{2} \pm
\frac{1}{2} \sqrt{(\widetilde{z}_u-\widetilde{z}_l)^2+4J^2}.
\end{align}
If $\widetilde{\gamma}_u = -\widetilde{\gamma}_l$ there is balanced loss and gain, and the system can exhibit coherent perfect absorption. 

The S-matrix of the empty resonator can be obtained directly from Eq. (\ref{eq:Smatrix}). Assuming equal system-bath couplings, and equal dampings $\gamma_u=\gamma_l=\gamma$, the indirect and direct transmission transmission matrix elements are ($t_{du}=t_{dl}=t_d$)
\begin{align}
\label{eq:tdti}
t_d &= \frac{-i 2\pi \kappa^2(\omega-\omega_a + i\gamma/2)}
{(\omega-\omega_1+i\gamma/2)(\omega-\omega_2+i\gamma/2)},
\\ \nonumber
t_i &= \frac{-i 2\pi \kappa^2 J}
{(\omega-\omega_1+i\gamma/2)(\omega-\omega_2+i\gamma/2)}.
\end{align}
Recall from Sec. \ref{sec:system} that $\omega_1$ ($\omega_2$) are the low (high) frequency cavity modes, and $J$ is the coupling between resonators. The zero in the numerator of $t_d$ at $\overline{z} = \omega_a-i\gamma/2$ is an antiresonance, or antimode, in the relevant transmission pathways. These equations may be used to characterise the resonators (as discussed in Supplemental Information~\ref{ap:estimation} and Sec. \ref{sec:Eres} of the main text).

\end{document}